\documentclass[journal,twoside,web]{ieeecolor}
\usepackage{generic}
\usepackage{cite}
\usepackage[table]{xcolor}
\usepackage{amsmath,amssymb,amsfonts}
\usepackage{algorithmic}
\usepackage{graphicx}
\usepackage[percent]{overpic}
\usepackage{textcomp}
\def\BibTeX{{\rm B\kern-.05em{\sc i\kern-.025em b}\kern-.08em
    T\kern-.1667em\lower.7ex\hbox{E}\kern-.125emX}}
\begin{document}
\title{Synchronization and multi-cluster capabilities of oscillatory networks with adaptive coupling}
\author{Petro~Feketa, Alexander~Schaum, and Thomas~Meurer,~\IEEEmembership{Senior~Member,~IEEE}
\thanks{This document contains a pre-print version of the paper conditionally accepted for the publication at the IEEE Transactions on Automatic Control. Cite this article as: P. Feketa, A. Schaum, and T. Meurer, ''Synchronization and multi-cluster capabilities of oscillatory networks with adaptive coupling'', \emph{IEEE Transactions on Automatic Control}, 2020.} \thanks{Manuscript submitted June 14, 2019. This work has been supported by the Deutsche Forschungsgemeinschaft (DFG) within the research unit FOR 2093: Memristive devices for neuronal systems (subproject C3: Synchronization of Memristively Coupled Oscillator Networks -- Theory and Emulation).}
\thanks{Petro Feketa is with Chair of Automatic Control, Kiel University, Germany (e-mail: pf@tf.uni-kiel.de). }
\thanks{Alexander Schaum is with Chair of Automatic Control, Kiel University, Germany (e-mail: alsc@tf.uni-kiel.de). }
\thanks{Thomas Meurer is with Chair of Automatic Control, Kiel University, Germany (e-mail: tm@tf.uni-kiel.de). }}

\newtheorem{theorem}{Theorem}
\newtheorem{lemma}{Lemma}
\newtheorem{corollary}{Corollary}
\newtheorem{proposition}{Proposition}
\newtheorem{example}{Example}
\newtheorem{definition}{Definition}
\newtheorem{remark}{Remark}

\newcommand\nred{\cellcolor{red!10}}
\newcommand\ngrey{\cellcolor{black!10}}
\newcommand\ngrn{\cellcolor{green!10}}

\definecolor{myAs}{rgb}{0.85,0.325,0.098}
\definecolor{myA}{rgb}{0.301,0.745,0.933}

\maketitle

\begin{abstract}
We prove the existence of a multi-dimensional non-trivial invariant toroidal manifold for the Kuramoto network with adaptive coupling. The constructed invariant manifold corresponds to the multi-cluster behavior of the oscillators phases. Contrary to the static coupling, the adaptive coupling strengths exhibit quasiperiodic oscillations preserving zero phase-difference within clusters. The derived sufficient conditions for the existence of the invariant manifold provide a trade-off between the natural frequencies of the oscillators, coupling plasticity parameters, and the interconnection structure of the network. Furthermore, we study the robustness of the invariant manifold with respect to the perturbations of the interconnection topology and establish structural and quantitative  constraints on the perturbation adjacency matrix preserving the invariant manifold. Additionally, we demonstrate the application of the new results to the problem of interconnection topology design which consists in endowing the desired multi-cluster behavior to the network by controlling its interconnection structure.
\end{abstract}


\section{Introduction}

Experimental studies of living organisms evince the decisive role of phase synchronization phenomena for many cognitive processes~\cite{rodriguez1999perception, fries2001modulation}. For example, the synchronization of oscillatory phases between different brain regions supports the interaction between working and long-term memory~\cite{fell2011role}. Complex neuro-inspired networks may exhibit partial synchronization and multi-cluster behavior due to their embedded adaptive capabilities. In particular, synaptic plasticity provides the basis for most models of learning, memory and development in real neural circuits~\cite{abbott2000synaptic}. {\color{black}In chemical systems, the network adaptation in response to its dynamics has been reported in \cite{Jain543}, where the reaction rates adapt dynamically depending on the state of the system. Another examples of activity-dependent adaptation can be found in social systems \cite{gross2007adaptive} that model opinion dynamics and beliefs propagation. Besides the studies \cite{rodriguez1999perception, fries2001modulation,fell2011role}, the phenomena of partial synchronization and multi-clustering are of practical relevance for neurophysiological systems \cite{rattenborg2000behavioral} and distributed power generation \cite{balaguer2010control}.}

A simple yet dynamically rich Kuramoto model proved to be an appropriate paradigm for synchronization phenomena~\cite{acebron2005kuramoto, dorfler2014synchronization}. Moreover, the Kuramoto-like oscillator networks may serve as adequate mathematical models for neural activity processes~\cite{cabral2011role, menara2019framework}.
The studies on full and partial synchronization as well as multi-cluster behavior of oscillator networks have attracted many researchers from statistical physics, nonlinear dynamical systems, and control communities. In particular, important results on frequency and phase synchronization can be found in monographs~\cite{boccaletti2008synchronized,pikovsky2003synchronization,strogatz2004sync}. Control theoretic approaches to the study of synchronization phenomena can be found in~\cite{DBSIAM, jadbabaie2004stability, chopra2009exponential, lin2007state, scardovi2007synchronization, schmidt2012frequency, sarlette2009synchronization}. A large amount of results on partial synchronization exploit certain network symmetries for the existence of cluster-synchronized states, see, e.g.,~\cite{PhysRevE626332, POGROMSKY200265, stewart, pst, Sorrentinoe1501737}.

Recent results on full sunchronization of adaptive and non-adaptive networks can be found in \cite{2018arXiv180807263Z, jafarpour2018synchronization, ha2016synchronization, ha2018emergent, gushchin2016phase}. In particular, \cite{ha2016synchronization, ha2018emergent} provide sufficient conditions in terms of inital states for the full phase and frequency synchronization of adaptive networks with particular learning rules. Frequency synchronization of adaptive Kuramoto networks for some particular number of clusters is studied in \cite{gushchin2016phase}.

In \cite{bronski2017stability}, a relation between adaptive and non-adaptive Kuramoto networks is provided and stability properties of fixed points of adaptive Kuramoto networks are studied. Also, numerical simulations with periodically oscillating behavior of the coupling strength and partially synchronized oscillator phases are presented there. Our paper provides a rigorous mathematical reasoning for the existence of the mentioned oscillating behavior. We also show that, in general, these oscillations are not periodic in time since they are generated by trajectories (which may be, for example, quasiperiodic) on a surface of a multi-dimensional invariant torus. 

Phase multi-clustering is characterized by a partition of the network nodes into subsets where the {\color{black}nodes'} phases evolve identically within each subset. Such subsets are called \emph{clusters}. For the case of static coupling we refer the reader to the series of papers \cite{menara2019framework, menara2019exact, menara2019stability,8263710} and \cite{4961065,4961435}, where the emergence and stability of multi-cluster behaviors for oscillatory networks with static coupling are studied. In \cite{5718119} the equlibriums which correspond to the multi-clustering in phase models with state-dependent adaptive coupling are studied. In our paper, we derive sufficient conditions for the existence {\color{black}and partial exponential stability}  of a multi-dimensional invariant toroidal manifold which corresponds to the multi-cluster behavior of the Kuramoto network with adaptive coupling. To the best of our knowledge, this is the first paper that proves the existence of such manifold. By this we extend {\color{black}the invariance} results of \cite{menara2019stability} to the case of adaptive networks and complement the results of \cite{5718119} by studying more complex limit behaviors of the system, namely invariant tori. 

Contrary to the notion of clustering used in this paper, in \cite{berner2018multi, berner2019self}, the authors study conditions for \emph{frequency clustering} in adaptive networks of identical Kuramoto oscillators and study possible arrangement of phases within every cluster. Existence criteria for multi-cluster solutions where different clusters correspond to different frequencies and their explicit form are presented.

Despite the discussed results, the global behavior of oscillatory networks is still far away from being fully understood. For example, in \cite{PhysRevLett.112.144103}, the emergence of a complex behavior that is characterized by the co-existence of regularly evolving clusters and irregularly oscillating nodes in Kuramoto networks with static coupling has been demonstrated. In the case of adaptive coupling, the phenomenon of self-organized emergence of hierarchical multilayered structures and chimera states is reported in \cite{PhysRevE.96.062211}.

The paper is organized as follows. In Section~\ref{sec2}, we recall some theoretical results from qualitative theory of {\color{black}nonlinear} dynamical systems defined in the product of a torus and the Euclidean space~\cite{Sam1, mitr} and demonstrate their relation to the synchronization analysis of oscillatory networks. In Section~\ref{sec3}, we prove the existence of an invariant toroidal manifold for the network of Kuramoto oscillators with adaptive coupling which corresponds to the multi-cluster behavior of the network. The proof of the main result is based on the perturbation theory of smooth invariant tori~\cite{SAMOILENKO19973121}. In Section~\ref{sec4}, we study the robustness of the constructed invariant manifold with respect to the perturbations of the interconnection topology and propose a design methodology which enriches the network to exhibit the desired multi-cluster behavior by adjusting its interconnection structure. Concluding remarks in Section~\ref{sec5} complete the paper. 

\subsection*{Notation} Let $\mathbb N${\color{black}, $\mathbb R$,} and $\mathbb R_{>0}$ denote the sets of natural{\color{black}, real,} and positive real numbers, respectively. For given $n,m\in\mathbb N$ let $\mathbb R^n$ and $\mathcal T_m$ denote the $n$-dimensional Euclidean space and $m$-dimensional torus, respectively. {\color{black}One-dimensional torus $\mathcal T_1$ is the one-sphere (circle).} Let $f:\mathcal T_m \to \mathbb R^n$ be a function of the variable $\varphi=(\varphi_1,\ldots,\varphi_m)^\top\in\mathcal T_m$ which is continuous and $2\pi$-periodic with respect to each $\varphi_s$, $s=\overline{1,m}$. By $C(\mathcal T_m)$ we denote the space of all such functions $f$ equipped with the norm
\begin{equation*}\label{Eq:zeronorm}
|f|_0 = \max_{\varphi \in \mathcal T_m} \left\|f(\varphi)\right\|,
\end{equation*}
where $\left\|\cdot\right\|$ denotes the Euclidean norm in $\mathbb R^n$, i.e., $\left\|f(\varphi)\right\|=\sum_{i=1}^n |f_i(\varphi)|^2$, $|f_i(\varphi)|$ stands for the absolute value of the $i$-th component of $f$ evaluated at $\varphi$. By $C^1(\mathcal T^m)$ we denote the subspace of $C(\mathcal T_m)$ with every $f\in C^1(\mathcal T_m)$ having a continuous partial derivative with respect to each $\varphi_s$, $s=\overline{1,m}$ and
\begin{equation*}\label{Eq:onenorm}
|f|_1 = \max \left\{|f|_0, \left|\frac{\partial f}{\partial \varphi_1}\right|_0, \ldots, \left|\frac{\partial f}{\partial \varphi_m}\right|_0 \right\}.
\end{equation*}
For any square matrix A we introduce the norm $\left\|A\right\|=\max\limits_{\left\|x\right\|=1}\left\|Ax\right\|$. For any rectangular matrix $B=(b_{ij})$ we use the Frobenius norm $\left\|B\right\|_F=\sum\limits_{i}\sum\limits_j{|b_{ij}|}$. A matrix $B:\mathcal T_m \to \mathbb R^{n\times m}$ is assumed to be of class $C(\mathcal T_m)$ if all its entries belongs to $C(\mathcal T_m)$. 

\section{Invariant tori of systems defined in $\mathcal T_m \times \mathbb R^n$}\label{sec2}

\subsection{Systems defined in $\mathcal T_m \times \mathbb R^n$}
We consider a system of ordinary differential equations defined in the direct product of the $m$-dimensional torus $\mathcal{T}_m$ and the $n$-dimensional Euclidean space $\mathbb{R}^n$
\begin{equation}
\label{MainEq1}
\begin{split}
\frac{d\varphi}{dt}=a(\varphi),\quad
\frac{dx}{dt}&=F(\varphi,x),
\end{split}
\end{equation}
where $\varphi=(\varphi_{1},\ldots,\varphi_{m})^{\top}\in\mathcal T_{m}$,
$x=(x_{1},\ldots,x_{n})^{\top}\in\mathbb R^{n}$, $a\in C(\mathcal T_{m})$, function $F:\mathcal T_m\times \mathbb R^n \to \mathbb R^n$ is continuous with respect to both arguments, and $F(\cdot,x) \in C(\mathcal T_m)$ for every fixed $x\in\mathbb R^n$.
We assume that there exists a positive constant $L>0$ such that for all $
\varphi{'},\varphi{''}\in\mathcal T_{m}$ it holds that 
\begin{equation}
\label{Lip}
\left\|a(\varphi{''})-a(\varphi{'})\right\|\leq
L\left\|\varphi{''}-\varphi{'}\right\|.
\end{equation}
Condition \eqref{Lip} guarantees that the system
\begin{equation}\label{DS}
\frac{d\varphi}{dt}=a(\varphi)
\end{equation}
generates a dynamical system on $\mathcal T_{m}$, which we shall denote by $\varphi_{t}(\varphi)$.

For any initial value $x^0 \in \mathbb R^n$ we denote by $x=x(t,\varphi,x^0)$ a solution to the Cauchy problem
\begin{equation}
\label{MainEq2}\begin{split}
\frac{dx}{dt}=F(\varphi_t(\varphi),x),
\quad x(0)=x^0 
\end{split}
\end{equation}
that depends on $\varphi\in\mathcal T_m$ as a parameter.

\begin{definition}[Invariant set]
A set $\mathcal S\subset \mathbb R^n\times \mathcal T_m$ is called an invariant set of \eqref{MainEq1}, if for all $(x^0,\varphi)\in\mathcal S$ and for all $t\geq 0$ the corresponding solution {\color{black}$(x(t, \varphi, x^0),\varphi_t(\varphi))\in\mathcal S$}. The invariant set of the system \eqref{MainEq1} is called nontrivial invariant toroidal manifold (or, simply, nontrivial invariant torus) if it is defined as 
$$\mathcal S_{u(\varphi)}=\left\{(x,\varphi)\in\mathbb R^n\times\mathcal T_m: x=u(\varphi),\, \varphi\in\mathcal T_m \right\}$$
 for some $u\in C(\mathcal T_m)$, $u(\varphi)\not\equiv 0$. If $u(\varphi)\equiv 0$ then the corresponding invariant set $$\mathcal S_0=\left\{(x,\varphi)\in\mathbb R^n\times\mathcal T_m: x=0,\, \varphi\in\mathcal T_m\right\},$$ is called trivial invariant toroidal manifold (or, simply, trivial invariant torus).
\end{definition}

\subsection{Relation to coupled Kuramoto oscillators}

For the purpose of illustration, let us demonstrate the relations between the existence of trivial and non-trivial invariant tori of systems of the type~\eqref{MainEq1} and synchronization capabilities of two coupled Kuramoto oscillators with static coupling. Consider the system
\begin{equation}\label{K2s}
\begin{array}{rcl}
\dot\theta_1 & = & w_1 + k\sin{(\theta_2-\theta_1)}, \\
\dot\theta_2 & = & w_2 + k\sin{(\theta_1-\theta_2)}, \\
\end{array}
\end{equation}
where the phases $\theta_1(t), \theta_2(t)\in\mathcal T_1$ for all $t\in\mathbb R$, coupling strength $k>0$ and natural frequencies $w_1, w_2>0$. Introducing the error $e=\theta_2-\theta_1$, we obtain 
\begin{equation}\label{K2e}
\begin{array}{rcl}
\dot\theta_1 & = & w_1 + k\sin{e}, \\
\dot e & = & w_2-w_1 - 2k\sin{e}, \\
\end{array}
\end{equation}
which we treat now as a system defined in $\mathcal T_1\times \mathbb R$ with $\theta_1\in\mathcal T_1$, $e\in(-\pi,\pi]$. The existence of a non-trivial invariant toroidal manifold
$$\mathcal S_d=\left\{(e,\theta_1)\in \mathbb R \times \mathcal T_1: e=d, \theta_1\in\mathcal T_1\right\}$$
for some constant $d\in\mathbb R$ corresponds to {\color{black}the} frequency synchronization in the Kuramoto interconnection \eqref{K2s}. Indeed, from the second equation of \eqref{K2e} we get
\begin{equation}\label{dest}
\begin{split}
0=w_2-w_1-2k\sin{d} \quad &\Rightarrow \quad \sin{d}=\frac{w_2-w_1}{2k} \\ &\Rightarrow \quad d=\arcsin{\frac{w_2-w_1}{2k}}.
\end{split}
\end{equation}
Then, from the first equation of \eqref{K2e},
\begin{equation*}
\dot \theta_1 = w_1+\frac{w_2-w_1}{2}=\frac{w_1+w_2}{2}
\end{equation*}
meaning that the states $\theta_1$ and $\theta_2$ will oscillate with the frequency that is the average of natural frequencies. Also, constant $d$ from \eqref{dest} can be used to conclude a practical synchronization of the phases: a larger coupling strength $k$ leads to a smaller $|d|$. In case of different natural frequencies, {\color{black}i.e.}, $w_1\not = w_2$, $d$ can be only non-zero meaning that system \eqref{K2e} does not posses a trivial invariant torus and phase synchronization in system \eqref{K2s} is not possible. This recovers a known result that the network of non-identical Kuramoto oscillators cannot be synchronized with static coupling~\cite{schmidt2012frequency}.

Finally, for the case of identical oscillators, {\color{black}i.e.}, $w_1=w_2=w$, system \eqref{K2e} possesses the trivial invariant torus
$$\mathcal S_0 = \left\{(e,\theta_1)\in\mathbb R\times\mathcal T_1: e\equiv 0, \theta_1\in\mathcal T_1 \right\}$$
that corresponds to the phase synchronization of oscillators with frequency $w$ and the non-trivial invariant torus   
$$\mathcal S_\pi = \left\{(e,\theta_1)\in\mathbb R\times\mathcal T_1: e\equiv \pi, \theta_1\in\mathcal T_1 \right\}$$
that corresponds to the anti-synchronization of phases.

\subsection{Auxiliary results from linear theory}
In this subsection, we briefly recall some results from the mathematical theory of multi-frequency oscillations that we require in the following sections. For a more comprehensive exposition, we refer the reader to the monographs~\cite{Sam1, mitr}. We shall specifically use the results for {\color{black}the} so-called linear extensions of dynamical systems on a torus
\begin{equation}
\label{Eq:LinearExt}
\begin{split}
\frac{d\varphi}{dt}=a(\varphi),\quad
\frac{dx}{dt}&=P(\varphi)x+f(\varphi)
\end{split}
\end{equation}
with $P,f\in C(\mathcal T_m)$. Let $\Omega_\tau^t(\varphi)$ denote the fundamental matrix of the  homogeneous system
\begin{equation}\label{param}
\dot x = P(\varphi_t(\varphi)) x
\end{equation}
that depends on $\varphi$ as a parameter, such that $\Omega_\tau^\tau(\varphi)\equiv I$, where $I$ stands for the identity matrix.

\begin{theorem}[{\cite[Section~3.5]{Sam1}}]\label{Thm:ExpTori}
Let $$\left\|\Omega_0^t(\varphi)\right\|\leq K e^{-\gamma t}$$
for all $t\geq 0$, $\varphi \in\mathcal T_m$ and some positive constants $K$ and $\gamma$ independent of $\varphi$. Then, for any $f\in C(\mathcal T_m)$ system~\eqref{Eq:LinearExt} has an exponentially stable\footnote{Here, the local exponential stability is meant, i.e., the trajectories of \eqref{Eq:LinearExt} starting in a vicinity of the invariant manifold do not travel far away from the manifold and converge exponentially to the manifold when $t\to\infty$.} invariant toroidal manifold $x=u(\varphi)$, $\varphi\in\mathcal T_m$ with
$$
u(\varphi)=\int\limits_{-\infty}^0 \Omega_\tau^0(\varphi)f(\varphi_\tau(\varphi))d\tau
$$
and $$|u|_0\leq K \gamma^{-1}|f|_0.$$
\end{theorem}

Let $S(\varphi)$ be a matrix from the space $C(\mathcal T_m)$.

\begin{definition}\label{GreenF}
A function
\begin{equation*}
G_0(\tau,\varphi)=\begin{cases}
\Omega^0_\tau(\varphi) S(\varphi_\tau(\varphi)), \qquad\qquad\,\, \tau\leq 0,\\
-\Omega^0_\tau(\varphi)(I-S(\varphi_\tau(\varphi))), \quad \tau> 0
\end{cases}
\end{equation*}
is called a Green-Samoilenko function of the system
\begin{equation*}
\dot \varphi = a(\varphi), \quad \dot x = P(\varphi)x,
\end{equation*}
if the integral $\int\limits_{-\infty}^\infty \left\|G_0(\tau,\varphi) \right\|d\tau$ is bounded uniformly with respect to $\varphi$.
\end{definition}

In particular, the Green-Samoilenko function exists if for any $\varphi\in\mathcal T_m$ system \eqref{param} is exponentially dichotomous on the entire real axis $\mathbb R$~\cite{mitr}. In the case of a constant matrix $P(\varphi)\equiv P$, this corresponds to the absence of purely imaginary eigenvalues. 

\begin{theorem}[{\cite[Theorem~11.1]{mitr}}]\label{Thm:SmoothTori}
Let $a, A \in C^1(\mathcal T_m)$ and for any
$f\in C^1(\mathcal T_m)$
the Green-Samoilenko function $G_0$ satisfies the inequality
\begin{equation}\label{Eq:CondForSmoothness}
\left| G_0(\tau,\varphi) f(\varphi_{\tau}(\varphi)) \right|_1 \leq K e^{-\gamma |\tau|} \left|f\right|_1, \quad \tau \in \mathbb R
\end{equation}
for some positive constants $K$ and $\gamma$ independent of $\varphi$. Then, system \eqref{Eq:LinearExt} has an invariant toroidal manifold $x=u(\varphi)$, $\varphi\in\mathcal T_m$ with $u\in C^1(\mathcal T_m)$ and 
$$
|u|_1\leq 2K\gamma^{-1}|f|_1.
$$
\end{theorem}

\section{Multi-clustering in adaptive networks of Kuramoto oscillators}\label{sec3}

In this section, we derive sufficient conditions for the existence of invariant toroidal manifolds for the network of Kuramoto oscillators with adaptive coupling. These manifolds correspond to the multi-cluster behavior of the phases.
 
\subsection{Main result}

Let ${\color{black}\mathbb G} = (\mathcal V, \mathcal E)$ be the directed graph representing the network of oscillators, where $\mathcal V=\{1,\ldots,N\}$ and $\mathcal E \subseteq \mathcal V \times \mathcal V$ represent the oscillators and their interconnection edges, respectively. Let $A = [a_{ij}]_{i,j=\overline{1,N}}$ be the adjacency matrix of ${\color{black}\mathbb G}$, where $a_{ij}=1$ if the edge $(i,j)\in\mathcal E$, and $a_{ij}=0$ when $(i,j)\not\in\mathcal E$. We assume that the graph does not have self-loops, i.e., $a_{ii}=0$ for all $i=\overline{1,N}$. The dynamics of the network is
\begin{equation}\label{main1}
\begin{array}{rcll}
\dot \theta_i & = & w_i+\sum\limits_{j=1}^N a_{ij} k_{ij} \sin(\theta_j-\theta_i), & i=\overline{1,N}, \\
\dot k_{ij} &=& -\gamma k_{ij} + \mu\Gamma(\theta_j-\theta_i), & i,j=\overline{1,N},
\end{array}
\end{equation}
where $w_i\in\mathbb R$ and $\theta_i(t)\in {\color{black}\mathcal T_1}$ denote the natural frequency and the phase of the $i$-th oscillator. The dynamics of the coupling strength $k_{ij}(t)\in\mathbb R$ is defined by positive parameters $\mu, \gamma \in \mathbb R_{>0}$ and $\Gamma\in C^1(\mathcal T_m)$ with $|\Gamma|_1=\delta\in\mathbb R_{>0}$.

A network exhibits cluster synchronization when the oscillators can be partitioned so that the phases of the oscillators in each cluster evolve identically. This type of behavior corresponds to the existence of an invariant toroidal manifold of system \eqref{main1}. Moreover, for the case of adaptive coupling this manifold is nontrivial which complicates the derivation of the conditions for its existence. 

Let $\mathcal P = \{\mathcal P_1,\ldots, \mathcal P_m\}$, with $m>1$, be a partition of $\mathcal V$, where $\cup_{i=1}^{m}\mathcal P_i =\mathcal V$ and $\mathcal P_i \cap \mathcal P_j = \varnothing$ if $i\not=j$. For a given partition $\mathcal P$ let us collect all non-zero entries of the adjacency matrix $A$ which correspond to the intra-cluster links and inter-cluster links into the sets $A_{in}$ and $A_{out}$, respectively. The cardinality of these sets
\begin{equation}\label{Eq:cardinality}
c_{in}=\mathtt{card}\{A_{in}\} \quad \text{and} \quad c_{out}=\mathtt{card}\{A_{out}\}
\end{equation}
characterize the interconnection structure of $\mathcal V$ with respect to the partition $\mathcal P$. Additionally, let
\begin{equation*}
w_{min}=\min\limits_{i=\overline{1,N}} |w_i| \quad \text{and}\quad w_{max}=\max\limits_{i=\overline{1,N}} |w_i|.
\end{equation*}

\begin{theorem}\label{thm_main}
Let for system \eqref{main1} and a given partition $\mathcal P$ the following conditions hold true:
\begin{itemize}
\item[(A1)] for any $s=\overline{1,m}$ and for any $i,j\in \mathcal P_s$ $$w_i=w_j;$$
\item[(A2)] for any $s,r=\overline{1,m}$, $s\not =r$ there exist constants $c_{sr}\in\mathbb N$ such that for any $i\in\mathcal P_s$ $$\sum\limits_{j\in\mathcal P_r} a_{ij}=c_{sr};$$
\item[(A3)] for $c_{max}:=\max\limits_{s=\overline{1,m}} \sum\limits_{r\not=s}c_{sr}$ it holds that
\begin{equation}\label{eqA31}
w_{min}-\mu\gamma^{-1}\delta c_{max}>0
\end{equation}
and
\begin{equation}\label{eqA32}
4\frac{\mu}{\gamma^2} \delta \sqrt{c_{out}} \sum\limits_{\substack{s,r=\overline{1,m} \\ s\not = r}}{c_{sr}} \frac{w_{max}+\mu\gamma^{-1}\delta c_{max}}{w_{min}-\mu\gamma^{-1}\delta c_{max}}<1.
\end{equation}
\end{itemize}
Then, system \eqref{main1} has an invariant toroidal manifold which corresponds to the $m$-cluster behavior defined by the partition $\mathcal P$.
\end{theorem}

\begin{remark}\label{rem1}
Conditions (A1)-(A3) allow for the following interpretation:
\begin{itemize}
\item (A1) requires the natural frequencies to be equal within every cluster.
\item (A2) requires that the number of links coming to every node
  within a given cluster $\mathcal P_s$ from other given cluster
  $\mathcal P_r$, $r\not=s$ is the same. The number of incoming links to the nodes of $\mathcal P_s$ from the cluster other than $\mathcal P_r$ may be different. Also, (A2) restricts only the number of links and does not require any symmetry of the corresponding adjacency matrix. It is worth to highlight that the intra-cluster couplings are generally not required for the emergence of multi-cluster behavior in the network. This type of behavior may result from a proper interaction of nodes with the nodes from other clusters.
\item (A3) establishes the relations between the natural frequencies of the oscillators, plasticity parameters $\mu, \gamma, \delta$ and the inter-cluster interconnection topology. For a given network of Kuramoto oscillators and a given partition $\mathcal P$, conditions (A3) can always be satisfied by choosing a sufficiently small plasticity parameter $\mu$. {\color{black} Also, the inequalities \eqref{eqA31} and \eqref{eqA32} are more likely to be satisfied for a large plasticity parameter $\gamma$ and a small number of inter-cluster connections, i.e, for small values of $c_{max}$ and $c_{out}$.}
\end{itemize}
\end{remark}

\begin{remark}
Conditions (A1) and (A2) have appeared previously in \cite{menara2019stability} in the context of non-adaptive Kuramoto networks. In~\cite{menara2019stability}, these conditions arise from an algebraic condition that has to be fulfilled in order to prove the existence of a trivial invariant torus. In this paper, we prove the existence of non-trivial invariant toroidal manifold which corresponds to the oscillating behavior of the coupling strength while preserving the zero phase-difference within clusters.
\end{remark}

\subsection{Proof of Theorem~\ref{thm_main}}

A scheme of the proof is as follows. First, we rearrange the variables in \eqref{main1} and make linear transformations of variables so that the system becomes defined in the product of an $m$-dimensional torus ($m$ is the number of clusters) and $\left(N-m+c_{in}+c_{out}\right)$-dimensional Euclidean space. Then, we decompose the problem of the existence of a non-trivial invariant torus into a similar problem but of a lower dimension \eqref{main4} and additional algebraic constraint \eqref{main5}. The invariant torus of \eqref{main4} is constructed as a limit of an infinite sequence of invariant toroidal manifolds of the corresponding auxiliary systems. Here we prove the existence of an  invariant manifold for every auxiliary system, the smoothness of the  invariant manifolds and uniform convergence of the sequence of these manifolds.

For every cluster $\mathcal P_s$, $s=\overline{1,m}$ let us pick an arbitrary oscillator $i_s \in \mathcal P_s$ and denote its phase and natural frequency by $\varphi_s:=\theta_{i_s}$ and $\bar w_s:=w_{i_s}$, respectively. For every oscillator $i\in\mathcal P_s$, let us introduce the relative difference $e_i=\theta_i-\varphi_s$. Then, from \eqref{main1}, we get
\begin{align}
\label{main2-torus}
\dot\varphi_s  =  &\bar w_s +
\sum\limits_{j\in\mathcal P_s}a_{i_sj} k_{i_sj}\sin{e_j} \\ 
\notag
&\quad+\sum\limits_{r\not=s}\sum\limits_{j\in\mathcal P_r} a_{i_sj}k_{i_sj}\sin(e_j+\varphi_r-\varphi_s),  \quad\, s=\overline{1,m},\\
\label{main2-errors}
\dot e_i  =  &w_i-\bar w_s \\
\notag
&+ \sum\limits_{j\in\mathcal P_s}\left[a_{ij} k_{ij}\sin(e_j-e_i)-a_{i_sj} k_{i_sj}\sin{e_j}\right] &~  \\
\notag
&+\sum\limits_{r\not=s}\sum\limits_{j\in\mathcal P_r} \left[a_{ij} k_{ij}\sin(e_j-e_i+\varphi_r-\varphi_s)\right. &~\\
\notag
&\qquad\qquad\quad\left.- a_{i_sj} k_{i_sj}\sin(e_j+\varphi_r-\varphi_s) \right]\\
\notag
&\qquad\qquad\qquad\qquad\qquad\qquad\forall i\in \mathcal P_s\setminus\{i_s\},\, s=\overline{1,m},\\[3mm]
\label{main2-inter}
\dot k_{ij}  = &-\gamma k_{ij} + \mu \Gamma(e_j-e_i+\varphi_r-\varphi_s) \\
\notag
&\qquad\qquad\qquad\quad\forall i\in\mathcal P_s,\,\, \forall j\in\mathcal P_r,\, s\not=r, s,r=\overline{1,m},\\[3mm]
\label{main2-intra}
\dot k_{ij}  = &-\gamma k_{ij} + \mu \Gamma(e_j-e_i)\qquad  \forall i,j\in\mathcal P_s,\, i\not=j,\, s=\overline{1,m}.
\end{align}
System \eqref{main2-torus}-\eqref{main2-intra} has the same number of equations as system \eqref{main1}. Equations \eqref{main2-torus} describe the dynamics of $m$ arbitrarily selected oscillators (one from every cluster). Equations \eqref{main2-errors} describe the error dynamics within each cluster. Equations \eqref{main2-inter} describe the dynamics of the coupling strength between nodes of different clusters. Finally, \eqref{main2-intra} describe the dynamics of the intra-cluster coupling strengths. Let us collect all $\varphi_i$, $i=\overline{1,m}$ into the vector $\varphi = (\varphi_1, \ldots, \varphi_m)^{\top}\in\mathcal T_m$ and all relative errors $e_i$, $i\in \mathcal P_s\setminus\{i_s\}$, $s=\overline{1,m}$ into the vector $e\in\mathbb R^{N-m}$. All inter- and intra-cluster coupling strengths we collect into the vectors $k^{inter}\in\mathbb R^{c_{out}}$ and $k^{intra}\in\mathbb R^{c_{in}}$, respectively, and $k=(k^{inter}, k^{intra})^\top$.
The multi-cluster behavior in network \eqref{main1} is possible if system \eqref{main2-torus}-\eqref{main2-intra} possesses an invariant toroidal manifold
\begin{equation}\label{main3}
e \equiv  0,\quad  k = u(\varphi), \quad \varphi\in\mathcal T_m
\end{equation}
for some $u\in C(\mathcal T_m)$.
The invariant manifold \eqref{main3} exists if there exists an invariant toroidal manifold $$k=u(\varphi),\quad \varphi\in\mathcal T_m$$ for the system
\begin{equation}\label{main4}
\begin{aligned}
\dot\varphi_s & = \bar w_s +\sum\limits_{r\not=s} \sum\limits_{j\in\mathcal P_r} a_{i_sj}k_{i_sj}\sin(\varphi_r-\varphi_s), \,\,\, s=\overline{1,m},\\
\dot k_{ij} & = -\gamma k_{ij} + \mu \Gamma(\varphi_r-\varphi_s) \\
 &\qquad\qquad\quad\,\,\, \forall i\in\mathcal P_s,\, \forall j\in\mathcal P_r,\, s\not=r,\, s,r=\overline{1,m}, \\
\dot k_{ij} &= -\gamma k_{ij} + \mu \Gamma(0) \qquad \forall i,j\in\mathcal P_s,\, i\not=j,\, s=\overline{1,m}.
\end{aligned}
\end{equation}
such that for all $\varphi\in\mathcal T_m$
\begin{equation}\label{main5}
\begin{array}{rcl}
0 & = & w_i-\bar w_s + \\ & & \sum\limits_{r\not=s} \sum\limits_{j\in\mathcal P_r} \left[ (a_{ij} u_{ij}(\varphi) - a_{i_sj} u_{i_sj}(\varphi))\sin(\varphi_r-\varphi_s) \right]\\ & & \forall i\in\mathcal P_s\setminus\{i_s\},\, s=\overline{1,m}.
\end{array}
\end{equation} 

From the last two equations of \eqref{main4} it follows that 
\begin{itemize}
\item the invariant toroidal manifold for the intra-cluster couplings exists and it is defined by $$k_{ij} = u^{(intra)}(\varphi)=\frac{\mu\Gamma(0)}{\gamma}$$
for all $i,j\in\mathcal P_s$, $i\not=j$, $s=\overline{1,m}$ since its dynamics is independent of $\varphi$. 
\item if the invariant toroidal manifold for the inter-cluster couplings exists, it has to be defined by functions which are the same for every link between two given clusters, {i.e., for all $i\in\mathcal P_s$, $j\in\mathcal P_r$, $s\not=r$, $s,r=\overline{1,m}$ functions $u_{ij}(\varphi)=u_{i_s i_r}(\varphi)$ for all $\varphi\in\mathcal T_m$ since the dynamics of $k_{ij}$ depends only on the dynamics of $\varphi_r$ and $\varphi_s$.}
\end{itemize}

The problem of the existence of a non-trivial invariant toroidal manifold of \eqref{main2-torus}-\eqref{main2-intra} is now reduced to the question of the existence of the invariant tori $k_{ij}=u_{ij}(\varphi)$, $\varphi\in\mathcal T_m$ of the system
\begin{equation}\label{main8}
\begin{array}{rcl}
\dot\varphi_s & = & \bar w_s +\sum\limits_{r\not=s} \sum\limits_{j\in\mathcal P_r} a_{i_sj}k_{i_sj}\sin(\varphi_r-\varphi_s), \quad s=\overline{1,m},\\
\dot k_{ij} & = & -\gamma k_{ij} + \mu \Gamma(\varphi_r-\varphi_s) \\
& &  \qquad\qquad\,\,\,\,\forall i\in\mathcal P_s,\, \forall j\in\mathcal P_r,\, s\not=r,\, s,r=\overline{1,m}
\end{array}
\end{equation} 
such that 
\begin{equation}\label{main9}
\begin{array}{rcl}
0 & = & w_i-\bar w_s \\ & &  + \sum\limits_{r\not=s} u_{i_s i_r}(\varphi) \sin(\varphi_r-\varphi_s)\sum\limits_{j\in\mathcal P_r} (a_{ij}-a_{i_sj}) \\ & & \forall i\in\mathcal P_s\setminus\{i_s\},\, s=\overline{1,m}.
\end{array}
\end{equation} 

In particular, the condition \eqref{main9} is fulfilled if for all $s,r=\overline{1,m}$, $s\not = r$ simultaneously
\begin{equation}\label{main6}
w_i=\bar w_s \quad \text{and}\quad  \sum\limits_{j\in\mathcal P_r} (a_{ij} - a_{i_sj})=0 \quad \text{for all}\quad i\in\mathcal P_s
\end{equation}
hold true. Conditions (A1) and (A2) imply \eqref{main6}. Condition \eqref{main6} exactly recovers the recently obtained sufficient condition for cluster synchronization for non-adaptive Kuramoto networks with static couplings~\cite{menara2019stability}:
\begin{itemize}
\item the equal natural frequencies within each cluster;
\item the number of links coming from the nodes of any other cluster to each node within a given cluster is the same.
\end{itemize}
Hence, the problem addressed in \cite{menara2019stability} may be interpreted as the problem of the existence of the trivial invariant torus $e\equiv 0$, $\varphi\in\mathcal T_m$. Proving the existence of the trivial torus for systems defined in $\mathcal T_m\times\mathbb R^{N-m}$ always reduces to solving the corresponding algebraic equation and does not require additional techniques from the qualitative theory of nonlinear dynamical systems. Our problem still requires proving the existence of a nontrivial torus of \eqref{main8} due to non-static coupling $k$.

System \eqref{main8} can be rewritten in the form
\begin{subequations}\label{general}
\begin{align}\label{generala}
\dot\varphi & = \bar w + B(A_{out},\varphi)k^{inter}, \\\label{generalb}
\dot k^{inter} & = -\gamma I k^{inter} + \mu \mathcal G(\varphi),
\end{align}
\end{subequations}
with $\bar w = (\bar w_1, \ldots, \bar w_m)^\top$, $c_{out}\times m$-dimensional matrix $B\in C^1(\mathcal T_m)$ and $c_{out}$-dimensional vector $\mathcal G\in C^1(\mathcal T_m)$. Non-zero entries of matrix $B$ are $a_{i_sj}\sin(\varphi_r-\varphi_s)$ with appropriate indices $r,s\in\{1,\ldots,m\}$, $j\in\mathcal P_r$. The number of such non-zero entries is $\textstyle\sum_{{s,r=\overline{1,m}, s\not = r}}{c_{sr}}$, i.e., the number of incoming links to the nodes $i_s$, $s=\overline{1,m}$ from all nodes of clusters $\mathcal P_r$, $r\not =s$. Hence, the Frobenius norm of $B$ can be estimated as
\begin{equation}\label{Eq:bounded}
\max\limits_{\varphi\in\mathcal T_m}\left\|B(A_{out},\varphi)\right\|_{F} \leq \sum\limits_{\substack{s,r=\overline{1,m} \\ s\not = r}}{c_{sr}}.
\end{equation}
Entries of $c_{out}$-dimensional vector function $\mathcal G\in C^1(\mathcal T_m)$ have a form $\Gamma(\varphi_r-\varphi_s)$ with appropriate indices $s,r=\overline{1,m}$, $s\not=r$ and, hence,
\begin{equation}\label{Eq:bounded2}
\left|\mathcal G(\varphi)\right|_1 \leq \delta\sqrt{c_{out}}.
\end{equation}

Since the right-hand side of \eqref{generala} depends on $k^{inter}$ it is not possible to apply directly Theorem~\ref{Thm:ExpTori} in order to construct the invariant toroidal manifold of \eqref{general}. For this purpose we use the method of successive approximations~\cite[Section~4.1]{Sam1}. We shall look for the invariant manifold of \eqref{general} as a limit of a sequence of manifolds
\begin{equation}\label{lmanifold}
\left\{k^{inter}=u^{(l)}(\varphi), \quad \varphi\in\mathcal T_m\right\}_{l=0}^{\infty},
\end{equation}
where $k^{inter}=u^{(l+1)}(\varphi)$, $\varphi\in\mathcal T_m$ is the invariant toroidal manifold of the system
\begin{subequations}\label{main10}
\begin{align}\label{main10a}
\dot\varphi & = \bar w + B(A_{out},\varphi)u^{(l)}(\varphi), \\ \label{main10b}
\dot k^{inter} & = -\gamma I k^{inter} + \mu \mathcal G(\varphi),
\end{align}
\end{subequations}
Later, we show that the smoothness of each manifold \eqref{lmanifold} is required to prove the convergence of the sequence of invariant tori (see Equation~\eqref{zkp1}). 

Let $u^{(0)}(\varphi)\equiv 0\in C^1(\mathcal T_m)$. For each step $l\geq 1$, \eqref{main10} is a linear extension of dynamical system on the torus of the form \eqref{Eq:LinearExt}. Since the matrix $-\gamma I$ is Hurwitz, the corresponding homogeneous (with respect to $k^{inter}$) system possesses the Green-Samoilenko function
\begin{equation}\label{myGreen}
G_0(\tau,\varphi)=\begin{cases}
0, \qquad\,\,\, \tau> 0,\\
e^{\gamma I \tau}, \quad \tau\leq 0.
\end{cases}
\end{equation}
Denote by $\varphi^{(l)}_t(\varphi)$ a solution to \eqref{main10a} satisfying the initial condition $\varphi^{(l)}_0(\varphi)=\varphi$ for every $\varphi\in\mathcal T_m$. From Theorem~\ref{Thm:ExpTori}, the invariant toroidal manifold of \eqref{main10} is defined by
\begin{equation}\label{step-k}
u^{(l+1)}(\varphi) = \int\limits_{-\infty}^{0}e^{\gamma I \tau}\mu G(\varphi^{(l)}_\tau(\varphi))d\tau
\end{equation}
and
\begin{equation}
|u^{(l+1)}|_0\leq \frac{1}{\gamma}|\mu G|_0 \leq \frac{\mu}{\gamma}\delta\sqrt{c_{out}}.
\end{equation}
Next, we show that this torus is smooth. This will be needed later to prove the convergence of the sequence $\{u^{(l)}\}_{l=0}^\infty$.
Condition \eqref{Eq:CondForSmoothness} from Theorem~\ref{Thm:SmoothTori} suggests that the smoothness properties of invariant tori depend on the smoothness of the corresponding Green-Samoilenko function $G_0$ and the behavior of trajectories $\varphi_t(\varphi)$ on the surface of the torus. Condition \eqref{eqA31} from (A3) implies that the right-hand side of \eqref{main10a} is separated from zero. This enables to estimate the partial derivatives of the trajectories on the torus according to
\begin{equation}\label{estTra}
\begin{split}
&\max\limits_{\substack{s=\overline{1,m} \\ q=\overline{1,m}}}\left|\frac{\partial}{\partial \varphi_s}\left[\varphi^{(l)}_{\tau}(\varphi)\right]_q \right|_0
= \max\limits_{\substack{s=\overline{1,m} \\ q=\overline{1,m}}}
\frac{\frac{d}{dt} \left[\varphi^{(l)}_{\tau}(\varphi)\right]_q}{\frac{d}{dt}\varphi_s} \\
&= \max\limits_{\substack{s=\overline{1,m} \\ q=\overline{1,m}}} \frac{\bar w_q +\sum\limits_{r\not=q} \sum\limits_{j\in\mathcal P_r} a_{i_qj}k_{i_qj}\sin\Big(\big[\varphi_\tau^{(l)}(\varphi)\big]_r-\big[\varphi_\tau^{(l)}(\varphi)\big]_q\Big)}{\bar w_s +\sum\limits_{r\not=s} \sum\limits_{j\in\mathcal P_r} a_{i_sj}k_{i_sj}\sin(\varphi_r-\varphi_s)} \\
&\leq \frac{w_{max}+\mu\gamma^{-1}\delta c_{max}}{w_{min}-\mu\gamma^{-1}\delta c_{max}},
\end{split}
\end{equation}
where $[\varphi^{(l)}_{\tau}(\varphi)]_j$ denotes the $j$-component of the vector $\varphi^{(l)}_{\tau}(\varphi)$.
From \eqref{myGreen}, since $G_0(\tau,\varphi)=0$ for $\tau>0$, condition \eqref{Eq:CondForSmoothness} holds true for $\tau>0$. For $\tau\leq 0$, using \eqref{estTra}, \eqref{Eq:bounded}, and \eqref{myGreen}, we have
\begin{equation}\label{estt}
\begin{split}
&\left| G_0(\tau,\varphi) \mu \mathcal G(\varphi^{(l)}_{\tau}(\varphi)) \right|_1
= \left| e^{\gamma I \tau} \mu \mathcal G(\varphi_{\tau}^{(l)}(\varphi)) \right|_1 \\
&\leq e^{\gamma \tau} \mu \left|\mathcal G(\varphi^{(l)}_{\tau}(\varphi)) \right|_1 \\
&\leq e^{\gamma \tau} \mu \delta \sqrt{c_{out}} \max\left\{1, 2\max\limits_{\substack{s=\overline{1,m} \\ j=\overline{1,m}}}\left|\frac{\partial}{\partial \varphi_s}\left[\varphi^{(l)}_{\tau}(\varphi)\right]_j \right|_0 \right\}\\
&\leq 2 e^{\gamma \tau} \mu \delta \sqrt{c_{out}}\frac{w_{max}+\mu\gamma^{-1}\delta c_{max}}{w_{min}-\mu\gamma^{-1}\delta c_{max}} \\
&\leq 2  \frac{w_{max}+\mu\gamma^{-1}\delta c_{max}}{w_{min}-\mu\gamma^{-1}\delta c_{max}} e^{\gamma \tau} \mu |\mathcal G|_1, \quad \tau \leq 0.
\end{split}
\end{equation}
Then, from Theorem~\ref{Thm:SmoothTori}, the function $u^{(l+1)}\in C^1(\mathcal T_m)$ is bounded together with its partial derivatives with respect to $\varphi$: 
\begin{equation}
\begin{split}
\left|u^{(l+1)}(\varphi)\right|_1 &\leq  4\frac{\mu}{\gamma} \frac{w_{max}+\mu\gamma^{-1}\delta c_{max}}{w_{min}-\mu\gamma^{-1}\delta c_{max}} |\mathcal G|_1 \\
&\leq  4\frac{\mu}{\gamma} \delta \sqrt{c_{out}} \frac{w_{max}+\mu\gamma^{-1}\delta c_{max}}{w_{min}-\mu\gamma^{-1}\delta c_{max}}
.
\end{split}
\end{equation}
Next, we show that
\begin{equation}
\lim\limits_{l\to\infty} u^{(l)}(\varphi)=u^{(inter)}(\varphi)
\end{equation}
uniformly with respect to $\varphi\in\mathcal T_m$. Since function $u^{(l+1)}(\varphi)$ defines the invariant torus of \eqref{main10} it satisfies the following partial differential equation
\begin{equation}\label{PDE2}
\begin{split}
\frac{\partial u^{(l+1)}(\varphi)}{\partial \varphi}\left( \bar w+B(A_{out},\varphi)u^{(l)}(\varphi)\right) = &-\gamma I u^{(l+1)}(\varphi) \\ &+\mu \mathcal G(\varphi).
\end{split}
\end{equation}
Analogously, 
\begin{equation}\label{PDE1}
\begin{split}
\frac{\partial u^{(l)}(\varphi)}{\partial \varphi}\left( \bar w+B(A_{out},\varphi)u^{(l-1)}(\varphi)\right) = &-\gamma I u^{(l)}(\varphi)\\ &+\mu \mathcal G(\varphi).
\end{split}
\end{equation}
Let $z^{(l+1)}(\varphi)=u^{(l+1)}(\varphi)-u^{(l)}(\varphi)$. Then, subtracting \eqref{PDE1} from \eqref{PDE2}, we have
\begin{equation}\label{tempnew}
\begin{split}
&\frac{\partial u^{(l+1)}(\varphi)}{\partial \varphi}\left( \bar w+B(A_{out},\varphi)u^{(l)}(\varphi)\right) \\ - &\frac{\partial u^{(l)}(\varphi)}{\partial \varphi}\left( \bar w+B(A_{out},\varphi)u^{(l-1)}(\varphi)\right) = -\gamma I z^{(l+1)}(\varphi).
\end{split}
\end{equation}
By subtracting and adding $$\frac{\partial u^{(l)}(\varphi)}{\partial \varphi}\left( \bar w+B(A_{out},\varphi)u^{(l)}(\varphi)\right)$$ in the left-hand side of \eqref{tempnew}, we arrive at 
\begin{equation}\label{PDE3}
\begin{split}
\frac{\partial z^{(l+1)}(\varphi)}{\partial \varphi}\left( \bar w+B(A_{out},\varphi)u^{(l)}(\varphi)\right) = &-\gamma I z^{(l+1)}(\varphi)\\&-f^{(l)}(\varphi),
\end{split}
\end{equation}
where
\begin{equation*}
f^{(l)}(\varphi) = \frac{\partial u^{(l)}(\varphi)}{\partial\varphi} B(A_{out},\varphi) z^{(l)}(\varphi).
\end{equation*}
Hence $x=z^{(l+1)}(\varphi)$, $\varphi\in\mathcal T_m$ is the invariant torus of
\begin{equation}
\begin{array}{rcl}
\dot\varphi & = & \bar w+B(A_{out},\varphi)u^{(l)}(\varphi),\\
\dot x & = & -\gamma I x - f^{(l)}(\varphi)
\end{array}
\end{equation} 
and, from Theorem~\ref{Thm:ExpTori}, it exists and satisfies the estimate
\begin{equation}\label{zkp1}
\begin{split}
\left|z^{(l+1)}\right|_0 &\leq \frac{1}{\gamma} \max_{\varphi\in\mathcal T_m}\left\| \frac{\partial u^{(l)}(\varphi)}{\partial \varphi} B(A_{out},\varphi) z^{(l)}(\varphi)\right\| \\
&\leq \frac{1}{\gamma} \max_{\varphi\in\mathcal T_m}\left\| \frac{\partial u^{(l)}(\varphi)}{\partial \varphi} B(A_{out},\varphi)\right\|_{F}\big| z^{(l)}\big|_0 \\
&\leq \frac{1}{\gamma} \big|u^{k}\big|_1\max_{\varphi\in\mathcal T_m}\left\| B(A_{out},\varphi)\right\|_{F}\big| z^{(k)}\big|_0 \\
&\leq
4\frac{\mu}{\gamma^2} \delta \sqrt{c_{out}} \sum\limits_{\substack{s,r=\overline{1,m} \\ s\not = r}}\hspace{-3mm}{c_{sr}} \frac{w_{max}+\mu\gamma^{-1}\delta c_{max}}{w_{min}-\mu\gamma^{-1}\delta c_{max}}\big|z^{(k)}\big|_0.
\end{split}
\end{equation}
Hence, $\lim\limits_{k\to\infty}\left|z^{(l)}(\varphi)\right| =0$ and $\lim\limits_{l\to\infty} u^{(l)}(\varphi)=u^{(inter)}(\varphi)$ if (A3) holds:
\begin{equation}\label{Eq:conv}
4\frac{\mu}{\gamma^2} \delta \sqrt{c_{out}} \sum\limits_{\substack{s,r=\overline{1,m} \\ s\not = r}}{c_{sr}} \frac{w_{max}+\mu\gamma^{-1}\delta c_{max}}{w_{min}-\mu\gamma^{-1}\delta c_{max}}<1.
\end{equation}
Condition \eqref{Eq:conv} defines the interrelation between the natural frequencies of the oscillators, plasticity parameters and the sparsity of the inter-cluster interconnection topology.

Let us show that the constructed invariant manifold $k^{inter}=u^{(inter)}(\varphi)$, $\varphi\in\mathcal T_m$ is the invariant toroidal manifold of the system \eqref{general}. Let $\varphi^*_t(\varphi)$ be a solution to $$\dot \varphi = \bar w + B(A_{out},\varphi)u^{inter}(\varphi)$$ with $\varphi^*_0(\varphi)=\varphi$. Then, $u^{(inter)}$ can be written as
\begin{equation}
u^{(inter)}(\varphi) = \int_{-\infty}^{0}e^{\gamma I \tau}\mu \mathcal G(\varphi^{*}_\tau(\varphi))d\tau.
\end{equation}
Then, function $t \mapsto u^{(inter)}(\varphi^*_t(\varphi))$ for any $\varphi\in\mathcal T_m$ satisfies the equality
\begin{equation*}
\frac{d}{dt}u^{(inter)}(\varphi^*_t(\varphi))=-\gamma I u^{(inter)}(\varphi^*_t(\varphi)) + \mu \mathcal G(\varphi^*_t(\varphi))
\end{equation*}
which proves that $k^{inter}=u^{(inter)}(\varphi)$, $\varphi\in\mathcal T_m$ is the invariant manifold of the system \eqref{general}.

Summarizing, we have proven that under conditions (A1), (A2), and (A3)
system \eqref{main2-torus}-\eqref{main2-intra} has the non-trivial invariant toroidal manifold
\begin{equation}\label{endtor}
\begin{split}
e\equiv 0, \quad k^{intra}&=\left(\frac{\mu\Gamma(0)}{\gamma}, \ldots, \frac{\mu\Gamma(0)}{\gamma}\right)^\top,\\  k^{inter}&=u^{(inter)}(\varphi), \quad \varphi\in\mathcal T_m,
\end{split}
\end{equation}
which corresponds to the multi-cluster synchronization of the Kuramoto network \eqref{main1} given by partition $\mathcal P$. This completes the proof.\hfill$\blacksquare$

\begin{remark}\label{remSt}{\color{black} Condition (A2) is the only structural requirement for the network in order to possess the desired invariant toroidal manifold. Hence, no graph connectivity is required in Theorem~\ref{thm_main}. Moreover, it can be easily seen that the graph may have no links at all. In the latter case, however, the corresponding invariant torus cannot be asymptotically stable and, therefore the multi-cluster behavior is possible if the initial intra-cluster errors are zero. In case, when there is at least one incoming inter-cluster coupling for every cluster, i.e., $c_{max}\geq 1$, the partial asymptotic stability of the invariant toroidal manifold can be concluded (i.e., stability with respect to part of the variables).}
Since the Green-Samoilenko function $G_0$ defined in \eqref{GreenF} allows for the estimate $\left\|G_0(\tau,\varphi)\right\|\leq e^{\gamma|\tau|}$ for $\tau \in \mathbb R$ uniformly with respect to $\varphi\in\mathcal T_m$, the constructed invariant toroidal manifold \eqref{endtor} is exponentially stable with respect to the coupling strength $k$.
\end{remark}

\begin{remark}
The conditions of Theorem~\ref{thm_main} are sufficient for the existence of an $m$-dimensional invariant toroidal manifold. They are not necessary since the derivation is based on the perturbation theory of smooth invariant tori that is conservative. Moreover, there may exist other non-trivial invariant tori than the ones expressed by the Green-Samoilenko function. In this context, it is of high interest to explore the potential  of differential dissipativity theory~\cite{8451900} for proving the existence and stability of invariant toroidal manifolds which represent a low-dimensional dominant behavior of high-dimensional nonlinear systems. {\color{black} Moreover, there may exist invariant submanifolds of the constructed invariant tori. These submanifolds and their stability properties will then particularize the mutual dynamic behavior of clusters and are of high interest for further research.}
\end{remark}

\begin{remark}
In the case of Hebbian learning rule~\cite{plasticityHebb}, i.e., $\Gamma(s)=\cos{(s)}$ and $m=2$ the conditions (A1), (A2) imply the existence of a non-trivial invariant toroidal manifold that is defined by the positive constant $\frac{\mu}{\gamma}$ for intra-cluster couplings and negative constant $-\frac{\mu}{\gamma}$ for inter-cluster couplings:
\begin{equation}\label{consttorus}
\begin{split}
e\equiv 0, \quad k^{intra}&=\left(\frac{\mu\Gamma(0)}{\gamma}, \ldots, \frac{\mu\Gamma(0)}{\gamma}\right)^\top, \\k^{inter}&=\left(\frac{\mu\Gamma(\pi)}{\gamma}, \ldots, \frac{\mu\Gamma(\pi)}{\gamma}\right)^\top, \quad \varphi\in\mathcal T_m.
\end{split}
\end{equation}
This can be easily checked by plugging \eqref{consttorus} into the equations \eqref{main2-torus}-\eqref{main2-intra}.
This type of invariant tori corresponds to the two-cluster behavior with the constant inter-cluster phase difference equal to $\pi$. From the view-point of frequency multi-clustering studied in~\cite{berner2018multi}, the discussed type of behavior corresponds to the one-cluster formation with \emph{anti-phase synchronous oscillators}.
\end{remark}
\color{black}

\subsection{Numerical example}
\begin{figure}[!ht]
\hspace{-5.2mm}
\begin{overpic}[width=0.27\textwidth]{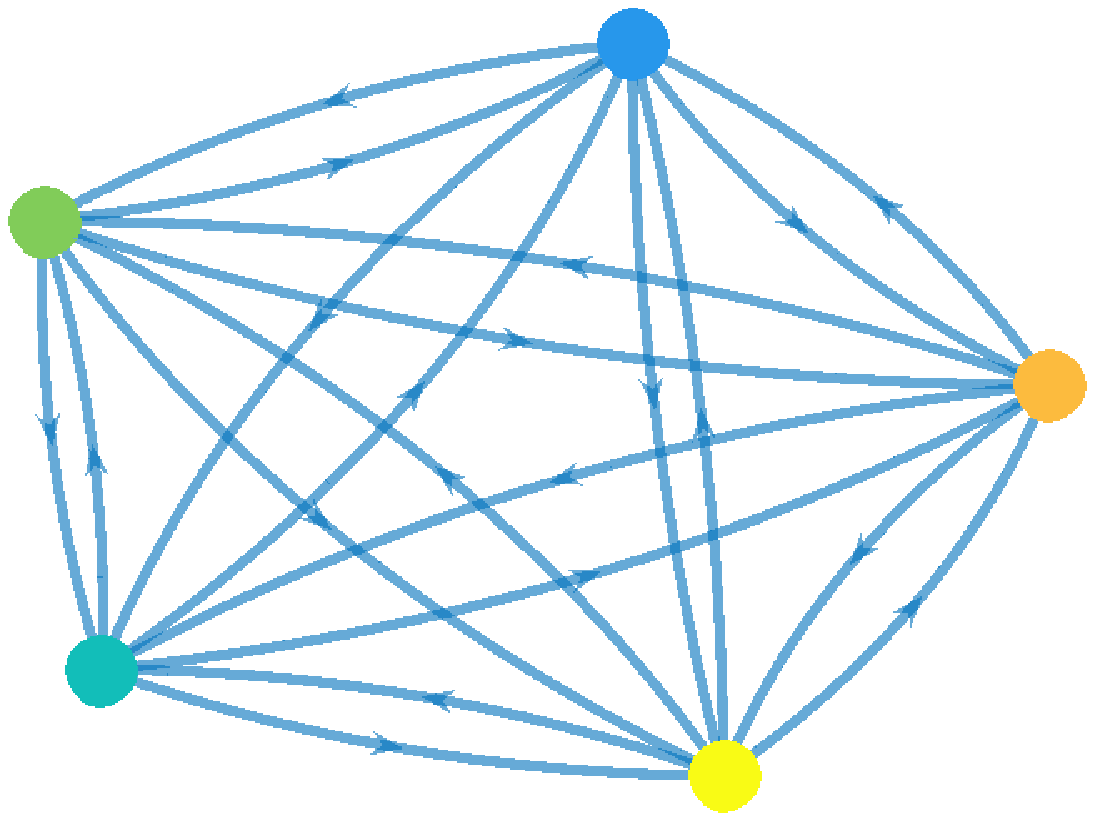}
 \put (57.8,60.3) {\tiny$\displaystyle 1$}
 \put (21.2,17.4) {\tiny$\displaystyle 2$}
 \put (17.5,48.2) {\tiny$\displaystyle 3$}
 \put (86.3,36.6) {\tiny$\displaystyle 4$}
 \put (64.0,10.0) {\tiny$\displaystyle 5$}
\end{overpic}\hspace{-5.8mm}
\begin{overpic}[width=0.27\textwidth]{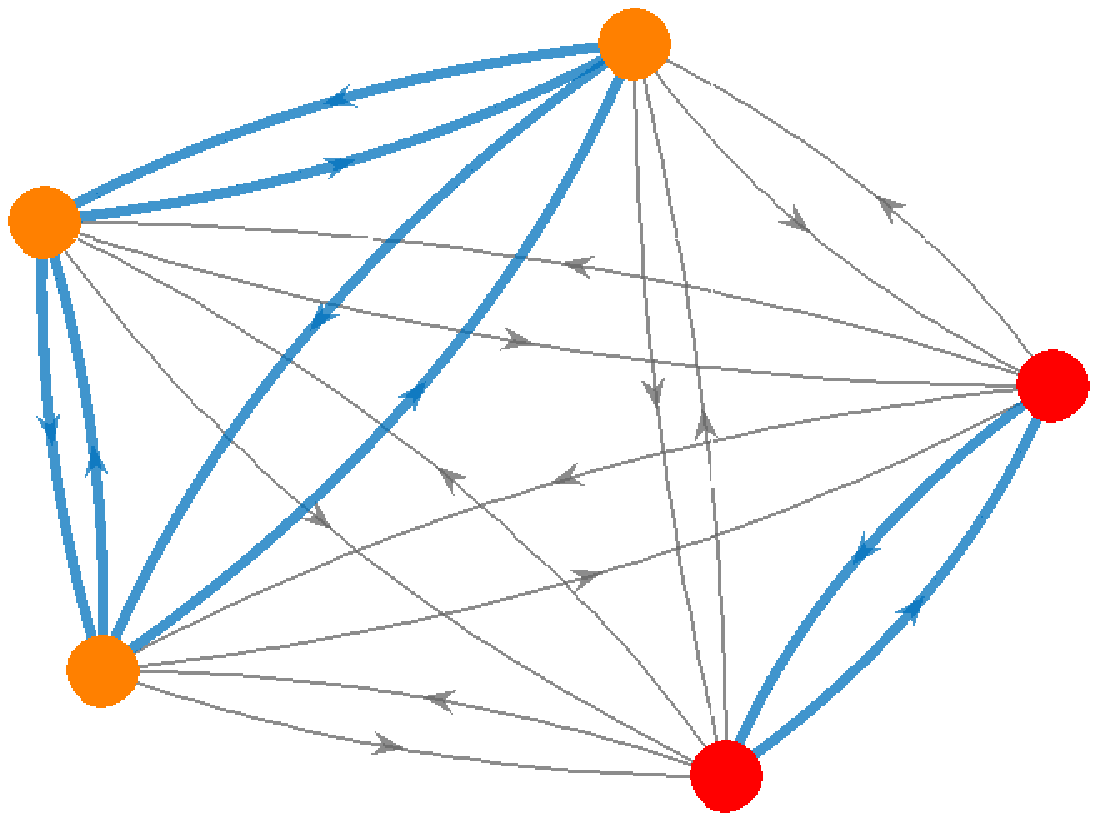}
 \put (57.6,60.3) {\tiny$\displaystyle 1$}
 \put (21.2,17.4) {\tiny$\displaystyle 2$}
 \put (17.0,48.2) {\tiny$\displaystyle 3$}
 \put (86.3,36.6) {\tiny$\displaystyle 4$}
 \put (64.0,10.0) {\tiny$\displaystyle 5$}
 \put (4,35) {\Large$\displaystyle \Rightarrow$}
\end{overpic}
\caption{{\color{black}A visualization of the graph $\mathbb G$ and coupling strengths} at the beginning (left figure) and at the end of simulation (right figure). Colors of the nodes represent their phases. Red nodes $1,2,3$ and orange nodes $4,5$ belong to two different clusters. For the right figure, blue connections denote intra-cluster links {\color{black}whose} coupling strengths converge to a constant value. Light-grey links correspond to the oscillating inter-cluster couplings {\color{black}whose} quasiperiodic trajectories approach the invariant manifold defined by \eqref{endtor} (see also Fig.~\ref{Fig:5-coupling} and Fig.~\ref{Fig:5-surface}).}
\label{Fig:5-ba}
\end{figure}

\begin{figure}[!ht]
\center
\begin{overpic}[width=0.5\textwidth]{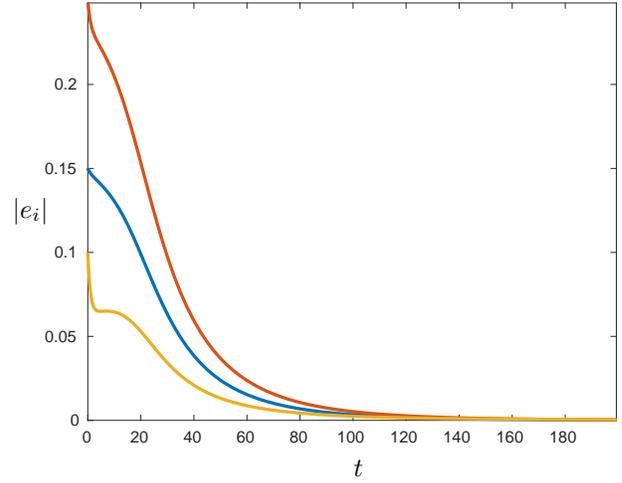}
 \put (52,0) {$\displaystyle t$}
 \put (2,38) {$\displaystyle |e_i|$}
\end{overpic}
\caption{Evolution of absolute values of the phase-errors $e_i$, $i=\overline{1,N}$ within clusters. Convergence of the errors to zero corresponds to the emergence of the multi-cluster partition of the network.}
\label{Fig:5-errors}
\end{figure}

\begin{figure}[!ht]
\center
\begin{overpic}[width=0.5\textwidth]{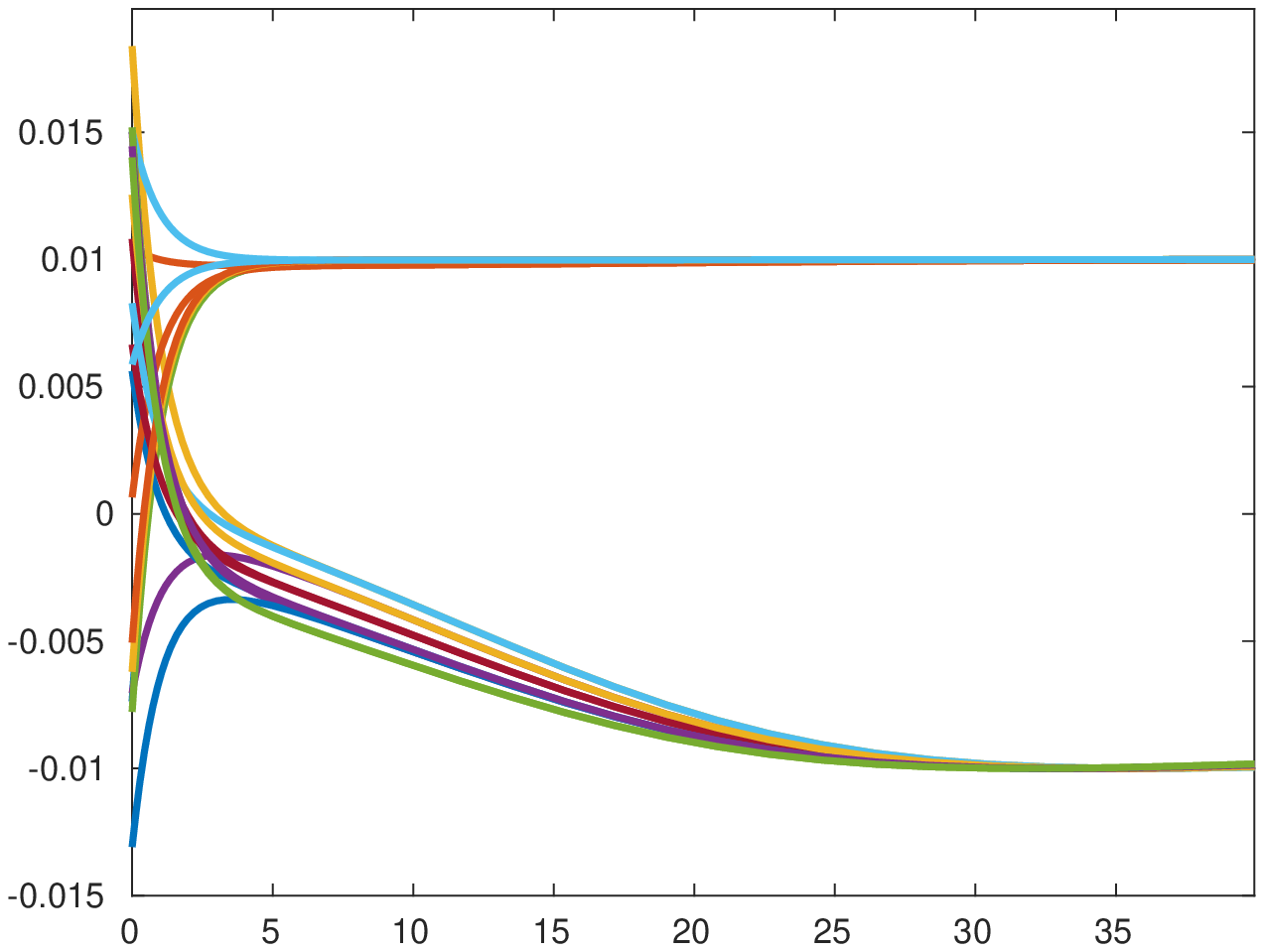}
 \put (52,0) {$\displaystyle t$}
 \put (2,33) {$\displaystyle k_{ij}$}
\end{overpic}
\begin{overpic}[width=0.5\textwidth]{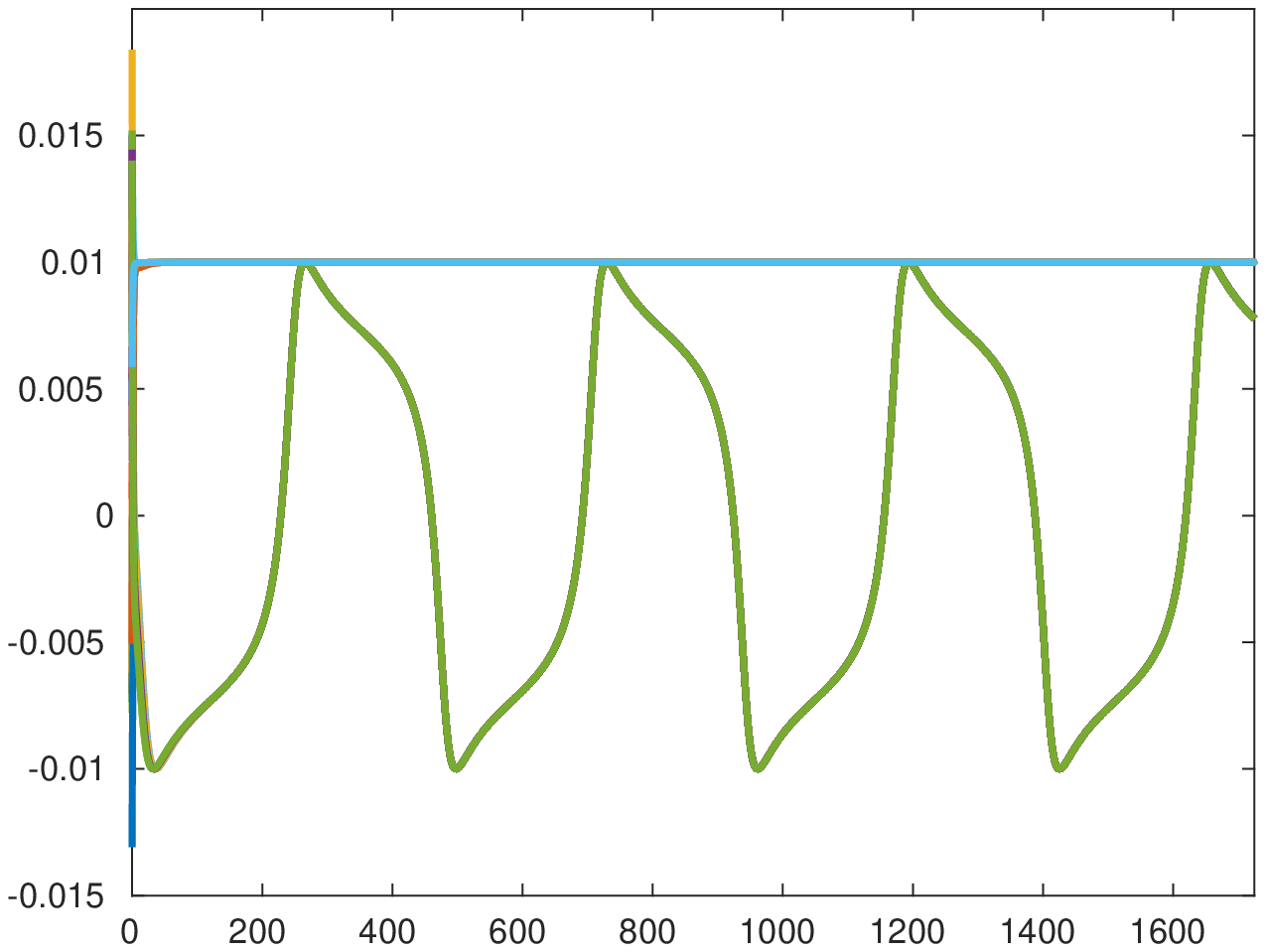}
 \put (52,0) {$\displaystyle t$}
 \put (2,33) {$\displaystyle k_{ij}$}
\end{overpic}
\caption{Evolution of coupling strengths $k_{ij}$ in different time scales: short-term (top figure) and long-term (bottom figure). The intra-cluster coupling strengths converge to the constant value $\frac{\mu}{\gamma}=0.01$ and the inter-cluster couplings exhibit quasiperiodic behavior and converge to the non-trivial invariant toroidal manifold \eqref{endtor}. The approximation of this manifold is depicted on Fig.~\ref{Fig:5-surface} and {\color{black}Fig.~}\ref{Fig:tori}.}
\label{Fig:5-coupling}
\end{figure}

\begin{figure}[!ht]
\begin{overpic}[width=0.5\textwidth]{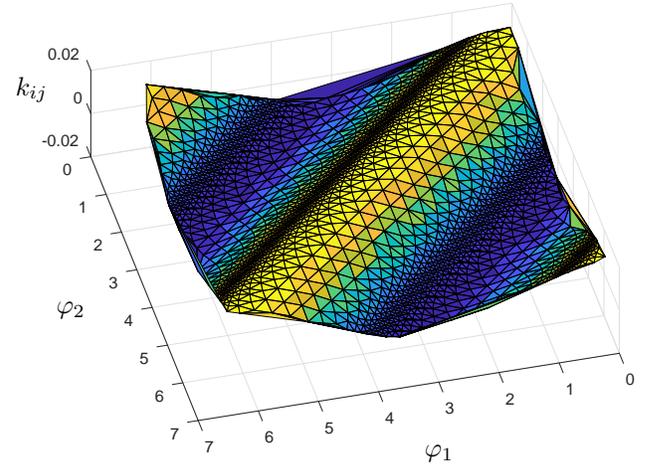}
 \put (62,3) {$\displaystyle \varphi_1$}
 \put (8,24) {$\displaystyle \varphi_2$}
 \put (2,56) {$\displaystyle k_{ij}$}
\end{overpic}
\caption{Approximation of the invariant manifold $k_{ij}= u_{ij}(\varphi)$, $\varphi\in \mathcal T_2$ for the inter-cluster coupling on torus map $[0,2\pi]\times [0,2\pi]$. The inter-cluster couplings $k_{ij}$ ''surf on the waves'' of the manifold exhibiting quasiperiodic behavior.}
\label{Fig:5-surface}
\end{figure}

\begin{figure}[!ht]
\begin{overpic}[width=0.45\textwidth]{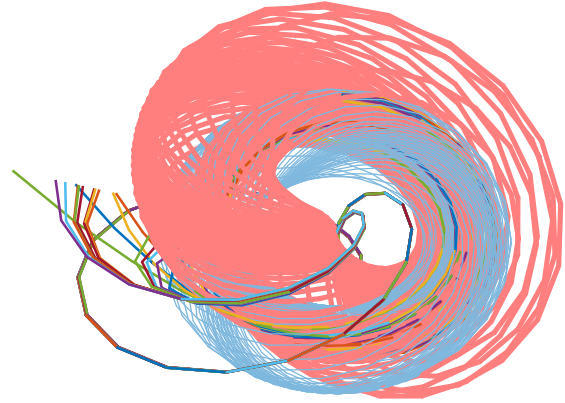}
\end{overpic}
\caption{Trajectories of $(k_{ij}, \varphi_1, \varphi_2)$ in the space $\mathbb R\times \mathcal T_2$ converge to the non-trivial torus \eqref{endtor} (in red) when $t\to\infty$. Trivial torus {\color{black}(which is not an invariant manifold for the considered system)} is plotted in light-blue for comparison.}
\label{Fig:tori}
\end{figure}

To illustrate the existence of invariant toroidal manifold and multi-cluster behavior of Kuramoto network we consider system \eqref{main1} with $N=5$ all-to-all connected nodes, natural frequencies $w = (0.5, 0.5, 0.5, \frac{\sqrt{2}}{3}, \frac{\sqrt{2}}{3})^\top$, plasticity parameters $\mu=0.01$, $\gamma_1=1$, Hebbian learning rule $\Gamma(s)=\cos(s)$, and the desired two-cluster partition $\mathcal P = \mathcal P_1 \cup \mathcal P_2 = \{1,2,3\}\cup \{4,5\}$. The natural frequencies satisfy the condition (A1). Condition (A2) is satisfied thanks to the all-to-all connections between nodes. We would like to note that, in general, a symmetry of the adjacency matrix is not necessary to fullfil (A2) (see also the example in Section IV and Remark 1 in~\cite{menara2019stability}). Directly calculating $c_{12}=2$, $c_{12}=3$, $c_{max}=3$, $c_{out}=12$, $\delta=1$, we check the condition (A3):
\begin{equation*}
w_{min}-\mu\gamma^{-1}\delta c_{max}= \frac{\sqrt{2}}{3}-\frac{0.01}{1}\cdot 3 \approx 0.4614 > 0
\end{equation*}
and
\begin{equation*}
\begin{split}
4\frac{\mu}{\gamma^2} \delta \sqrt{c_{out}} \sum\limits_{\substack{s,r=\overline{1,m} \\ s\not = r}}{c_{sr}} \frac{w_{max}+\mu\gamma^{-1}\delta c_{max}}{w_{min}-\mu\gamma^{-1}\delta c_{max}} \\ 
= 
4\frac{0.01}{1^2} \sqrt{12} (2+3) \frac{0.5+0.03}{\frac{\sqrt{2}}{3}-0.03}\approx 0.796 <1.
\end{split}
\end{equation*}  
Simulation results for initial phases $\theta^0 = (\frac{\pi}{2}, \frac{\pi}{2}+\frac{3}{20}, \frac{\pi}{2}+\frac{1}{4}, 0, -\frac{1}{10})^\top$ and random initial couplings $k_{ij}\in [-0.015, 0.015]$, $i,j=\overline{1,5}$, $i\not = j$ are presented on Fig{\color{black}s}.~\ref{Fig:5-ba}, \ref{Fig:5-errors}, \ref{Fig:5-coupling}, \ref{Fig:5-surface}, and {\color{black}\ref{Fig:tori}}. The relative phase-errors within clusters are chosen as $e_2=\theta_2-\varphi_1$, $e_3=\theta_3-\varphi_1$, and $e_5=\theta_5-\varphi_2$ with $\varphi_1:=\theta_1$ and $\varphi_2:=\theta_4$.

{\color{black}
\begin{remark}
A{\color{black}s it can be seen from Fig. \ref{Fig:5-coupling} and Fig.~\ref{Fig:5-surface}, the inter-cluster coupling strengths can take negative values. In general, the proposed approach for the construction of invariant manifolds does not necessarily require negative couplings (i.e., learning rule $\Gamma\in C^1(\mathcal T_m)$ can be chosen in such a way that the corresponding invariant torus is positive) and, therefore, our approach covers both cases of sign-changing and adaptive diffusive coupling.} 
\end{remark}
\begin{remark}\color{black}
Although the present paper is focused on the existence of invariant toroidal manifolds and provide only a limited result on partial exponential stability of tori w.r.t. {\color{black}coupling strengths (see Remark \ref{remSt}), the simulations suggest that the constructed invariant torus is asymptotically stable. We conjecture that the local exponential stability of multi-cluster formations heavily depends on the existence intra-cluster links, whose presence is not necessary for the existence of invariant toroidal manifolds (see condition (A2) and Remark~\ref{rem1}).}
\end{remark}
}

\color{black}

\section{Interconnection topology design for multi-cluster behavior of the network}\label{sec4}

In this section, we provide a Corollary from Theorem~\ref{thm_main} which quantifies the admissible perturbation of the adjacency matrix $A$ preserving the existence of the invariant toroidal manifold {\color{black}for the Kuramoto network with adaptive coupling}. On the other hand, the mentioned perturbation can be used as interconnection topology design tool ensuring the emergence of an invariant toroidal manifold. {\color{black}Recently, for oscillatory networks with static coupling, more comprehensive interconnection topology control techniques in form of certain optimal control problems have been proposed in \cite{menara2019framework} and \cite{8341829} for the purposes of multi-clustering and partial synchronization, respectively.}

Let the perturbation $\tilde A$ of the adjacency matrix $A$ be given by the entries $\tilde a_{ij}$ with $\tilde a_{ii}=0$ for all $i=\overline{1,N}$ and for any $i\not=j$
\begin{equation}
\tilde a_{ij} \in \begin{cases}
\{0,1\} \qquad \text{if}\quad a_{ij}=0,\\
\{-1,0\} \quad\, \text{if}\quad a_{ij}=1.
\end{cases}
\end{equation}
The values $\pm 1$ for the $\tilde a_{ij}$ correspond to removing the existing link or adding a new link the network, respectively. If $\tilde a_{ij}=0$ then no changes to the edge $(i,j)\in\mathcal E$ is made.

\begin{figure}[!ht]
\center
\begin{overpic}[width=0.5\textwidth]{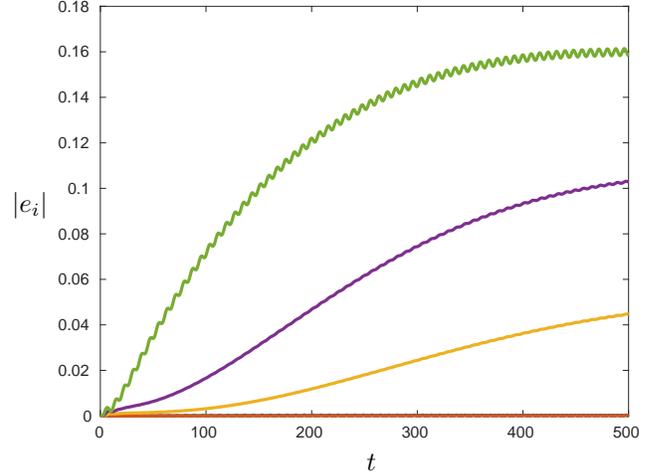}
 \put (52,0) {$\displaystyle t$}
 \put (0,38) {$\displaystyle |e_i|$}
\end{overpic}
\caption{Evolution of absolute values of the phase-errors $e_i$, $i=\overline{1,N}$ within clusters in case of adjacency matrix $A$. The errors $e_i$, $i=\overline{1,N}$ (which are initially set to zero) diverge from zero and converge to some new oscillating trajectories different from zero. Hence, the desired multi-cluster formation $\mathcal P$ is not achieved.}
\label{Fig:4-errors}
\end{figure}

\begin{corollary}
{Let for system \eqref{main1} with the adjacency matrix $A+\tilde A$ and given partition $\mathcal P$ the following conditions hold}
\begin{itemize}
\item[(A1)] for any $s=\overline{1,m}$ and for any $i,j\in \mathcal P_s$ $$w_i=w_j;$$
\item[(A2')] for any $s,r=\overline{1,m}$, $s\not =r$ there exist constants $\tilde c_{sr}\in\mathbb N$ such that for any $i\in\mathcal P_s$ $$\sum\limits_{j\in\mathcal P_r} a_{ij}+\tilde a_{ij}=\tilde c_{sr};$$
\item[(A3')] for $\tilde c_{max}:=\max\limits_{s=\overline{1,m}} \sum\limits_{r\not = s}\tilde c_{sr}$ it holds that
\begin{equation}\label{perttop1}
w_{min}-\mu\gamma^{-1}\delta \tilde c_{max}>0
\end{equation}
and
\begin{equation}\label{perttop2}
4\frac{\mu}{\gamma^2} \delta \sqrt{c_{out}+\tilde c_{out}} \sum\limits_{\substack{s,r=\overline{1,m} \\ s\not = r}}{\tilde c_{sr}} \frac{w_{max}+\mu\gamma^{-1}\delta \tilde c_{max}}{w_{min}-\mu\gamma^{-1}\delta \tilde c_{max}}<1,
\end{equation}
\end{itemize}
where $\tilde c_{out}$ is the sum of all entries of matrix $\tilde A$ which correspond to the inter-cluster links, i.e.,
\begin{equation}
\tilde c_{out}=\sum\limits_{s=1,m}\sum_{\substack{i\in \mathcal P_s \\ j\not\in\mathcal P_s}}\tilde a_{ij}.
\end{equation}
Then, the perturbed system has an invariant toroidal manifold which corresponds to the $m$-cluster behavior defined by the partition $\mathcal P$.\hfill$\blacksquare$
\end{corollary}
Conditions \eqref{perttop1}, \eqref{perttop2} suggest that by an appropriate choice of the perturbation matrix $\tilde A$ (e.g., removal of inter-cluster links by picking $\tilde A$ with $\tilde c_{out} \ll 0$), one may extend the set of admissible plasticity parameters guaranteeing the existence of an invariant toroidal manifold. Moreover, a network that is not capable of exhibiting a particular type of multi-cluster behavior can be restructured into a network that is capable of exhibiting the desired multi-cluster formation. In particular, this can be {\color{black}done} by proper {\color{black}adjustments} of the interconnection topology. We demonstrate it on the following network:

Consider a network of $N=7$ Kuramoto oscillators \eqref{main1} with adjacency matrix
\begin{equation*}
A=\left(
\begin{array}{ccc|cccc}
\ngrey 0 & \ngrey 1 & \ngrey 0 & 0 & \ngrn 1 & 0 & 0 \\
\ngrey 0 & \ngrey 0 & \ngrey 1 & 0 & 0 & 0 & \ngrn 1 \\
\ngrey 1 & \ngrey 0 & \ngrey 0 & \ngrn 1 & 0 & 0 & 0 \\ \hline
0 & \ngrn 1 & 0 & \ngrey 0 & \ngrey 1 & \ngrey 0 & \ngrey 0 \\
0 & \ngrn 1 & 0 & \ngrey 0 & \ngrey 0 & \ngrey 1 & \ngrey 0 \\
0 & 0 & \ngrn 1 & \ngrey 0 & \ngrey 0 & \ngrey 0 & \ngrey 1 \\
\nred 1 & 0 & \nred 1 & \ngrey 1 & \ngrey 0 & \ngrey 0 & \ngrey 0 \\
\end{array}
\right)
\end{equation*}
and natural frequencies $w = \left(\frac{1}{2}, \frac{1}{2}, \frac{1}{2}, \sqrt{\frac{4}{5}}, \sqrt{\frac{4}{5}}, \sqrt{\frac{4}{5}}, \sqrt{\frac{4}{5}}\right)^\top$, plasticity parameters $\gamma = 0.2$, $\mu = 0.001$, Hebbian learning rule $\Gamma(s)=\cos(s)$, and the desired two-cluster partition $\mathcal P = \{1,2,3\} \cup \{4,5,6,7\}$. Condition (A1) of Theorem~\ref{thm_main} is satisfied thanks to the choice of $w$. The adjacency matrix $A$ suggests that the network has exactly one incoming inter-cluster link for every node (green cells in $A$) except the node $7$ (red cells in $A$). Hence, the condition (A2) of Theorem~\ref{thm_main} is not satisfied. Fig.~\ref{Fig:4-errors} shows that even in the case of zero initial errors for intra-cluster phases ($e_i^0=0$, $i=\overline{1,N}$), zero inter-cluster initial coupling ($k_{15}^0=k_{27}^0=k_{34}^0=k_{42}^0=k_{52}^0=k_{63}^0=k_{71}^0=k_{73}^0=0$) and strong intra-cluster initial coupling ($k_{12}^0=k_{23}^0=k_{31}^0=k_{45}^0=k_{56}^0=k_{67}^0=k_{74}^0=1$), the phases of the oscillators do not exhibit two-cluster behavior. Choosing $\varphi_1:=\theta_1$, $\varphi_2:=\theta_4$, the intra-cluster errors $e_i$:
\begin{equation*}
\begin{split}
&e_2=\theta_2-\varphi_1, \quad e_3=\theta_3-\varphi_1, \\
e_5=\theta_5-\varphi_2, \quad &e_6=\theta_6-\varphi_2, \quad e_7=\theta_7-\varphi_2 
\end{split}
\end{equation*}
start exactly from zero and converge to some non-trivial invariant torus (see Fig.~\ref{Fig:4-errors}).

\begin{figure}[!ht]
\center
\begin{overpic}[width=0.5\textwidth]{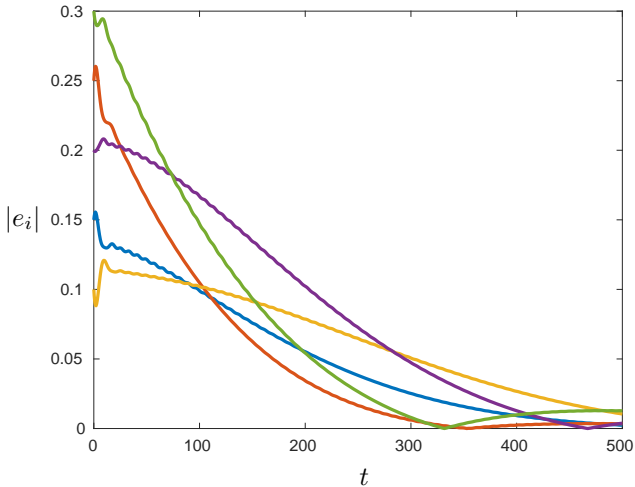}
 \put (52,0) {$\displaystyle t$}
 \put (0,38) {$\displaystyle |e_i|$}
\end{overpic}
\caption{Evolution of absolute values of the phase-errors $e_i$, $i=\overline{1,N}$ within clusters in case of the modified adjacency matrix $A^*$. All $e_i(t)\xrightarrow[]{t \to \infty} 0$, $i=\overline{1,N}$, which corresponds to the emergence of multi-cluster formation given by $\mathcal P$.}
\label{Fig:3-errors}
\end{figure}

Let us modify the network so that it satisfies the conditions of Theorem~\ref{thm_main}. For example, this can be made by removing the link between node $7$ and node $1$, i.e., $a_{17} := 0$. We denote the modified adjacency matrix by $A^*$. Now, every node in cluster $\mathcal P_1$ has exactly one incoming link from the nodes of cluster $\mathcal P_2$, and vice versa. The resulting interconnection topology satisfies the condition (A2) from Theorem~\ref{thm_main} and the network satisfies (A3) with characteristics $c_{out}=7$, $c_{max}=c_{12}=c_{21}=1$, $\delta = 1$. Indeed,
\begin{equation*}
w_{min}-\mu\gamma^{-1}\delta c_{max}= \frac{1}{2}-\frac{0.001}{0.2}=0.495>0
\end{equation*}
and
\begin{equation*}
\begin{split}
4\frac{\mu}{\gamma^2} \delta \sqrt{c_{out}} \sum\limits_{\substack{s,r=\overline{1,m} \\ s\not = r}}{c_{sr}} \frac{w_{max}+\mu\gamma^{-1}\delta c_{max}}{w_{min}-\mu\gamma^{-1}\delta c_{max}} \\ 
= 
4\frac{0.001}{0.2^2} \sqrt{7} (1+1) \frac{\sqrt{\frac{4}{5}}+0.005}{0.5-0.005}\approx 0.9615 <1.
\end{split}
\end{equation*}
All conditions of Theorem~\ref{thm_main} are satisfied. The corresponding intra-cluster errors converge to zero (see Fig.~\ref{Fig:3-errors}) even for the case of randomly picked initial coupling strengths $k_{ij}^0\in[-0.015, 0.015]$, which are allowed to be negative. The initial phases are taken as $$\theta^0 = \left(\frac{\pi}{2}, \frac{\pi}{2}+ \frac{3}{20}, \frac{\pi}{2}+ \frac{1}{4}, \frac{\pi}{3}- \frac{1}{10}, \frac{\pi}{3}- \frac{2}{10}, \frac{\pi}{3}- \frac{3}{10}\right)^\top.$$

\begin{figure}[!ht]
\center
\begin{overpic}[width=0.5\textwidth]{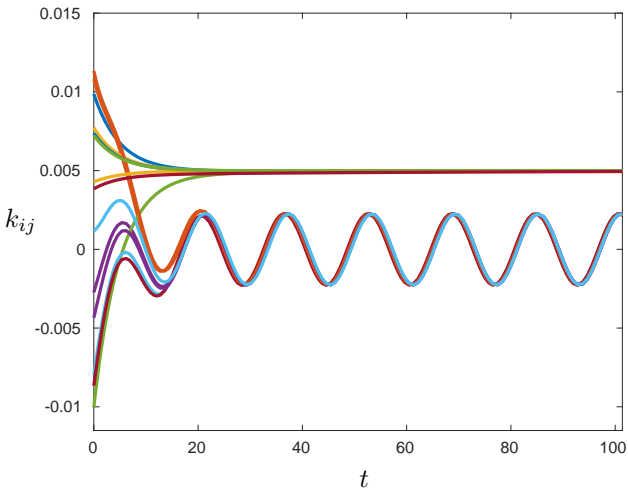}
 \put (52,0) {$\displaystyle t$}
 \put (0,38) {$\displaystyle k_{ij}$}
\end{overpic}
\caption{Evolution of coupling strengths $k_{ij}$. The intra-cluster coupling strengths converge to the constant value $\frac{\mu}{\gamma}=0.005$ and the inter-cluster couplings exhibit quasiperiodic behavior and converge to the non-trivial invariant toroidal manifold \eqref{endtor}.}
\label{Fig:3-coupling}
\end{figure}

The evolution of the coupling strengths of the system with the adjacency matrix $A^*$ is depicted on Fig.~\ref{Fig:3-coupling}.

\begin{figure}[!ht]
\hspace{-5mm}
\begin{overpic}[width=0.27\textwidth]{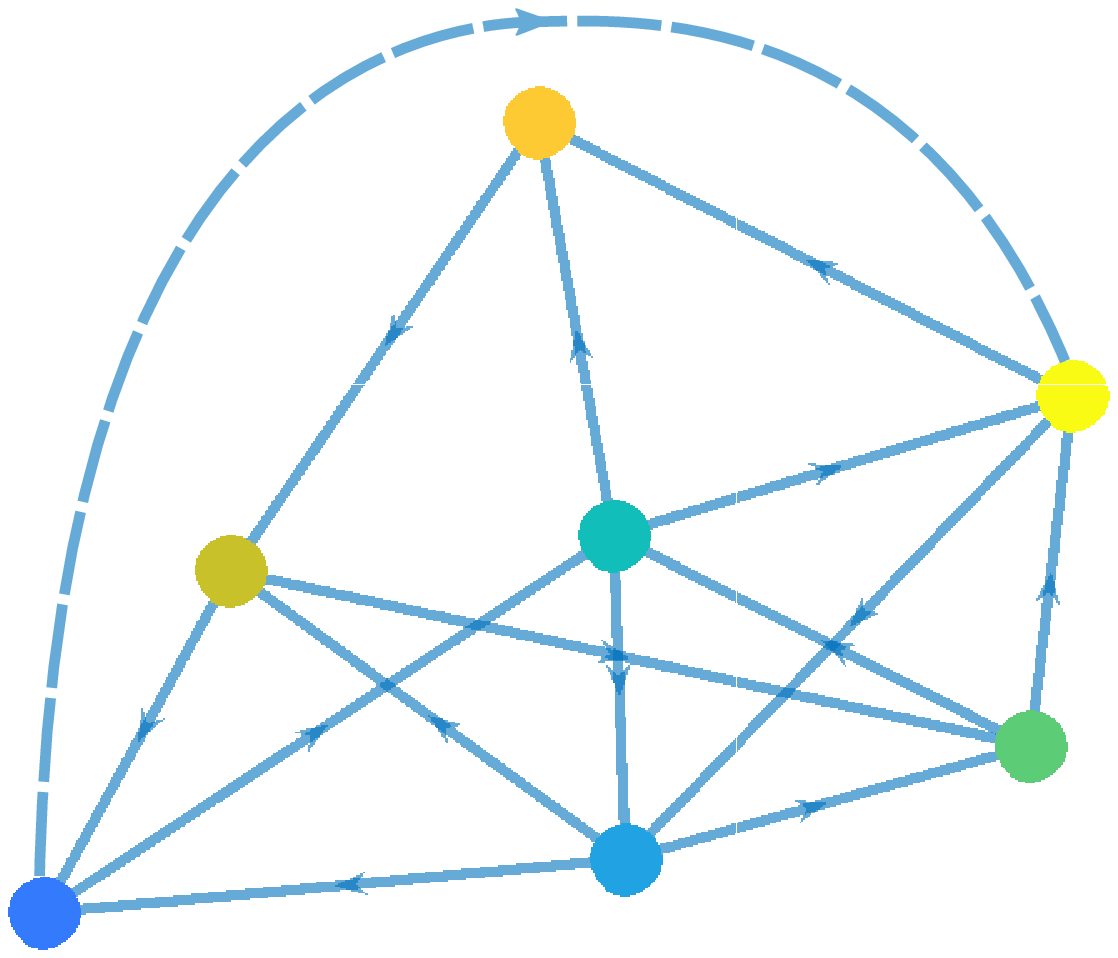}
 \put (15.3,10.5) {\tiny$\displaystyle 1$}
 \put (55.1,14.4) {\tiny$\displaystyle 2$}
 \put (54.4,36.6) {\tiny$\displaystyle 3$}
 \put (82.9,22.2) {\tiny$\displaystyle 4$}
 \put (28.2,34.0) {\tiny$\displaystyle 5$}
 \put (49,64.5) {\tiny$\displaystyle 6$}
 \put (85.8,45.9) {\tiny$\displaystyle 7$}
\end{overpic}\hspace{-6mm}
\begin{overpic}[width=0.27\textwidth]{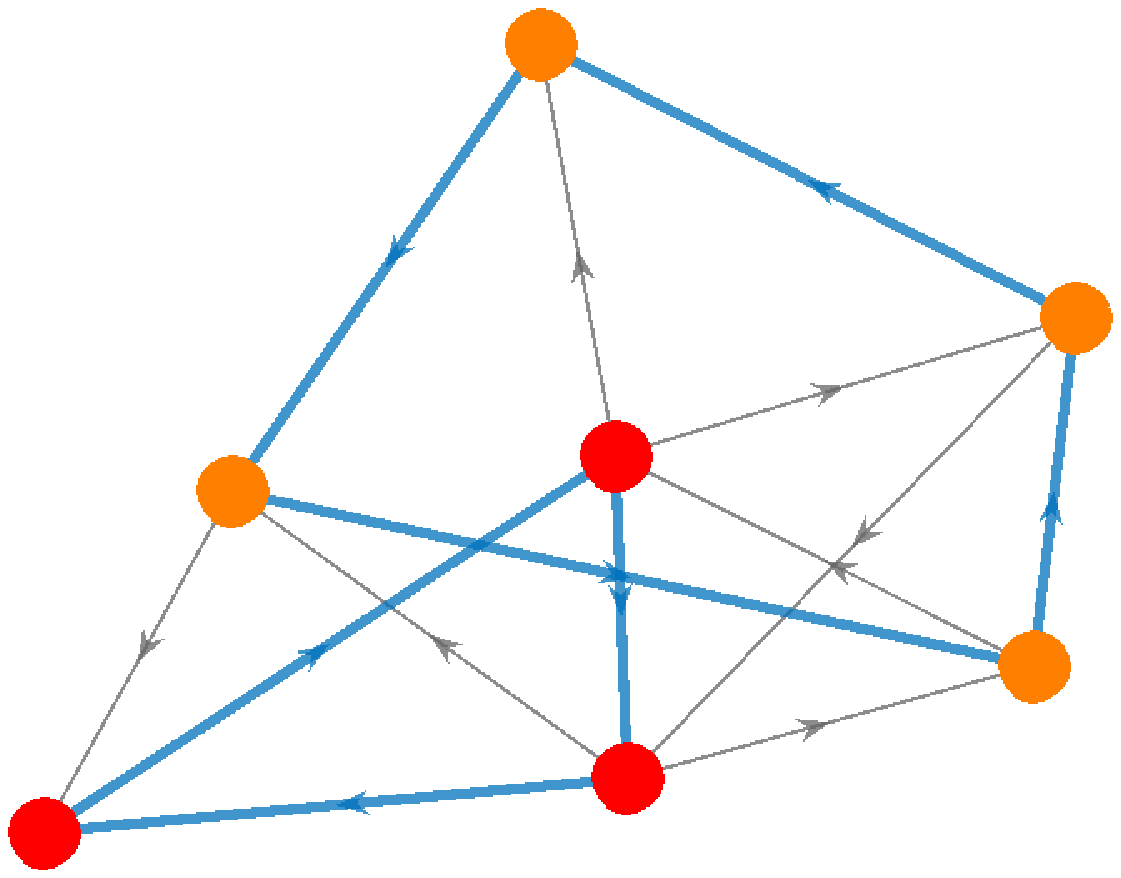}
 \put (15.3,10.5) {\tiny$\displaystyle 1$}
 \put (55.4,14.4) {\tiny$\displaystyle 2$}
 \put (54.6,36.4) {\tiny$\displaystyle 3$}
 \put (83.1,22.2) {\tiny$\displaystyle 4$}
 \put (28.2,34.0) {\tiny$\displaystyle 5$}
 \put (49.5,64.6) {\tiny$\displaystyle 6$}
 \put (85.99,45.9) {\tiny$\displaystyle 7$}
 \put (8,40) {\Large$\displaystyle \Rightarrow$}
\end{overpic}
\caption{Interconnection topology at the beginning (left figure) and at the end of simulation (right figure). Colors of the nodes represent their phases. Red nodes $1,2$ and $3$ and orange nodes $4,5,6$, and $7$ belong to two different clusters. {\color{black}Dashed link on the left figure corresponds to the inter-cluster coupling between nodes $1$ and $7$. This link will be removed at $t=500$.} For the right figure, blue connections denote intra-cluster links {\color{black}whose} coupling strengths converge to a constant value. Light-grey links correspond to the oscillating inter-cluster couplings {\color{black}whose} quasiperiodic trajectories approach the invariant manifold defined by \eqref{endtor} (see also Fig.~\ref{Fig:3-coupling}).}
\label{Fig:topology}
\end{figure}

\begin{figure}[!ht]
\center
\begin{overpic}[width=0.5\textwidth]{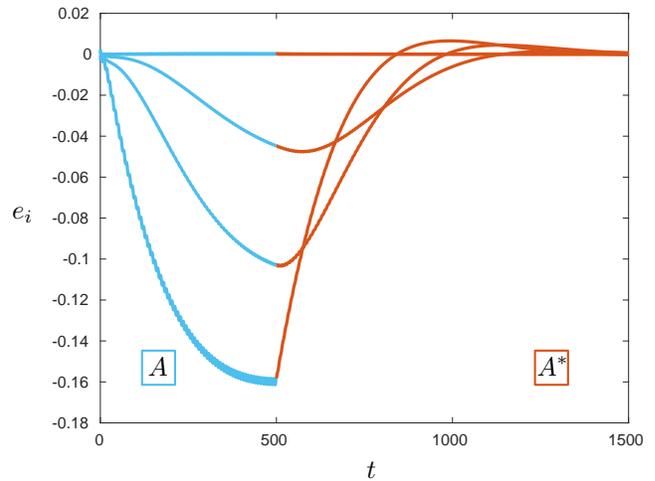}
 \put (52,0) {$\displaystyle t$}
 \put (0,38) {$\displaystyle e_{i}$}
 \put (20,15) {$\color{black}\displaystyle A$}
 \put (18.5,13.5) {\huge$\color{myA}\displaystyle \square$}
 \put (77,15) {$\color{black}\displaystyle A^*$}
 \put (76.1,13.5) {\huge$\color{myAs}\displaystyle \square$}
\end{overpic}
\caption{Evolution of the phase errors $e_i$, $i=\overline{1,N}$ within clusters in case of {\color{black}the} switching from $A$ (light blue curves) to $A^*$  (red curves) at $t=500$. The desired two-cluster behavior has been achieved by a proper change of the interconnection topology of the network.}
\label{Fig:6-control}
\end{figure}

Finally, Fig.~\ref{Fig:topology} and {\color{black}Fig.~\ref{Fig:6-control} depict the evolution of the interconnection topology and the} intra-cluster phase errors $e_i$, $i=\overline{1,N}$ for the scenario when the adjacency matrix $A$ is used for $t\in[0,500]$ and the interconnection {\color{black}structure} is switched according to $A^*$ at $t=500$. The desired multi-cluster formation has been achieved (i.e, $e_i(t)\xrightarrow[]{t \to \infty} 0$, $i=\overline{1,N}$) by a proper control of the interconnection topology of the network.

\section{Discussion and outlook}\label{sec5}

In this paper, we study the multi-clustering phenomenon in Kuramoto networks with adaptive coupling. This type of behavior corresponds to the existence of an invariant toroidal manifolds of the corresponding Kuramoto model. Sufficient conditions for the existence {\color{black}and partial exponential stability} of the mentioned invariant \mbox{manifolds} which inter-relate the interconnection topology of the network, natural frequencies of oscillators and plasticity parameters of adaptive couplings are established. Additionally, a methodology to achieve the capability of the desired multi-cluster behavior for the network by a proper design of its interconnection topology is proposed.

The derived sufficient conditions for the emergence of multi-cluster behavior require the same natural frequency for the oscillators within each cluster. This requirement also appears in~\cite{menara2019stability} for the case of Kuramoto network with static constant couplings. However, the plasticity of couplings in model \eqref{main1} offers flexible ways to achieve multi-clustering. In particular, condition~\eqref{main5} allows for a variety of different scenarios leading to multi-cluster behavior of the network by providing a relation between natural frequencies of oscillators, interconnection topology and quasiperiodic behavior of couplings. {\color{black}Hence,} sufficient conditions for the existence of other invariant toroidal manifolds than the one defined in \eqref{endtor} and their stability properties are of a great interest.  

\bibliography{references}

\begin{thebibliography}{10}
\providecommand{\url}[1]{#1}
\csname url@samestyle\endcsname
\providecommand{\newblock}{\relax}
\providecommand{\bibinfo}[2]{#2}
\providecommand{\BIBentrySTDinterwordspacing}{\spaceskip=0pt\relax}
\providecommand{\BIBentryALTinterwordstretchfactor}{4}
\providecommand{\BIBentryALTinterwordspacing}{\spaceskip=\fontdimen2\font plus
\BIBentryALTinterwordstretchfactor\fontdimen3\font minus
  \fontdimen4\font\relax}
\providecommand{\BIBforeignlanguage}[2]{{%
\expandafter\ifx\csname l@#1\endcsname\relax
\typeout{** WARNING: IEEEtran.bst: No hyphenation pattern has been}%
\typeout{** loaded for the language `#1'. Using the pattern for}%
\typeout{** the default language instead.}%
\else
\language=\csname l@#1\endcsname
\fi
#2}}
\providecommand{\BIBdecl}{\relax}
\BIBdecl

\bibitem{rodriguez1999perception}
E.~Rodriguez, N.~George, J.-P. Lachaux, J.~Martinerie, B.~Renault, and F.~J.
  Varela, ``Perception's shadow: long-distance synchronization of human brain
  activity,'' \emph{Nature}, vol. 397, no. 6718, p. 430, 1999.

\bibitem{fries2001modulation}
P.~Fries, J.~H. Reynolds, A.~E. Rorie, and R.~Desimone, ``Modulation of
  oscillatory neuronal synchronization by selective visual attention,''
  \emph{Science}, vol. 291, no. 5508, pp. 1560--1563, 2001.

\bibitem{fell2011role}
J.~Fell and N.~Axmacher, ``The role of phase synchronization in memory
  processes,'' \emph{Nature reviews neuroscience}, vol.~12, no.~2, p. 105,
  2011.

\bibitem{abbott2000synaptic}
L.~F. Abbott and S.~B. Nelson, ``Synaptic plasticity: taming the beast,''
  \emph{Nature neuroscience}, vol.~3, no. 11s, p. 1178, 2000.

\bibitem{Jain543}
\BIBentryALTinterwordspacing
S.~Jain and S.~Krishna, ``A model for the emergence of cooperation,
  interdependence, and structure in evolving networks,'' \emph{Proceedings of
  the National Academy of Sciences}, vol.~98, no.~2, pp. 543--547, 2001.
  [Online]. Available: \url{https://www.pnas.org/content/98/2/543}
\BIBentrySTDinterwordspacing

\bibitem{gross2007adaptive}
T.~Gross and B.~Blasius, ``Adaptive coevolutionary networks: a review,''
  \emph{Journal of the Royal Society Interface}, vol.~5, no.~20, pp. 259--271,
  2007.

\bibitem{rattenborg2000behavioral}
N.~C. Rattenborg, C.~Amlaner, and S.~Lima, ``Behavioral, neurophysiological and
  evolutionary perspectives on unihemispheric sleep,'' \emph{Neuroscience \&
  Biobehavioral Reviews}, vol.~24, no.~8, pp. 817--842, 2000.

\bibitem{balaguer2010control}
I.~J. Balaguer, Q.~Lei, S.~Yang, U.~Supatti, and F.~Z. Peng, ``Control for
  grid-connected and intentional islanding operations of distributed power
  generation,'' \emph{IEEE Transactions on Industrial Electronics}, vol.~58,
  no.~1, pp. 147--157, 2010.

\bibitem{acebron2005kuramoto}
J.~A. Acebr{\'o}n, L.~L. Bonilla, C.~J.~P. Vicente, F.~Ritort, and R.~Spigler,
  ``The {K}uramoto model: A simple paradigm for synchronization phenomena,''
  \emph{Reviews of modern physics}, vol.~77, no.~1, p. 137, 2005.

\bibitem{dorfler2014synchronization}
F.~D{\"o}rfler and F.~Bullo, ``Synchronization in complex networks of phase
  oscillators: A survey,'' \emph{Automatica}, vol.~50, no.~6, pp. 1539--1564,
  2014.

\bibitem{cabral2011role}
J.~Cabral, E.~Hugues, O.~Sporns, and G.~Deco, ``Role of local network
  oscillations in resting-state functional connectivity,'' \emph{Neuroimage},
  vol.~57, no.~1, pp. 130--139, 2011.

\bibitem{menara2019framework}
T.~Menara, G.~Baggio, D.~S. Bassett, and F.~Pasqualetti, ``A framework to
  control functional connectivity in the human brain,'' \emph{arXiv preprint
  arXiv:1904.08805}, 2019.

\bibitem{boccaletti2008synchronized}
S.~Boccaletti, ``The synchronized dynamics of complex systems,''
  \emph{Monograph series on nonlinear science and complexity}, vol.~6, pp.
  1--239, 2008.

\bibitem{pikovsky2003synchronization}
A.~Pikovsky, M.~Rosenblum, and J.~Kurths, \emph{Synchronization: a universal
  concept in nonlinear sciences}.\hskip 1em plus 0.5em minus 0.4em\relax
  Cambridge University Press, 2003, vol.~12.

\bibitem{strogatz2004sync}
S.~Strogatz, \emph{Sync: The emerging science of spontaneous order}.\hskip 1em
  plus 0.5em minus 0.4em\relax Penguin UK, 2004.

\bibitem{DBSIAM}
F.~D{\"o}rfler and F.~Bullo, ``On the critical coupling for {K}uramoto
  oscillators,'' \emph{SIAM Journal on Applied Dynamical Systems}, vol.~10,
  no.~3, pp. 1070--1099, 2011.

\bibitem{jadbabaie2004stability}
A.~Jadbabaie, N.~Motee, and M.~Barahona, ``On the stability of the {K}uramoto
  model of coupled nonlinear oscillators,'' in \emph{Proceedings of the 2004
  American Control Conference}, vol.~5.\hskip 1em plus 0.5em minus 0.4em\relax
  IEEE, 2004, pp. 4296--4301.

\bibitem{chopra2009exponential}
N.~Chopra and M.~W. Spong, ``On exponential synchronization of {K}uramoto
  oscillators,'' \emph{IEEE Transactions on Automatic Control}, vol.~54, no.~2,
  pp. 353--357, 2009.

\bibitem{lin2007state}
Z.~Lin, B.~Francis, and M.~Maggiore, ``State agreement for continuous-time
  coupled nonlinear systems,'' \emph{SIAM Journal on Control and Optimization},
  vol.~46, no.~1, pp. 288--307, 2007.

\bibitem{scardovi2007synchronization}
L.~Scardovi, A.~Sarlette, and R.~Sepulchre, ``Synchronization and balancing on
  the n-torus,'' \emph{Systems \& Control Letters}, vol.~56, no.~5, pp.
  335--341, 2007.

\bibitem{schmidt2012frequency}
G.~S. Schmidt, A.~Papachristodoulou, U.~M{\"u}nz, and F.~Allg{\"o}wer,
  ``Frequency synchronization and phase agreement in {K}uramoto oscillator
  networks with delays,'' \emph{Automatica}, vol.~48, no.~12, pp. 3008--3017,
  2012.

\bibitem{sarlette2009synchronization}
A.~{Sarlette} and R.~{Sepulchre},
  ``\BIBforeignlanguage{English}{{Synchronization on the circle.}}'' in
  \emph{\BIBforeignlanguage{English}{{The complexity of dynamical systems. A
  multi-disciplinary perspective}}}.\hskip 1em plus 0.5em minus 0.4em\relax
  Weinheim: Wiley-VCH, 2011, pp. 213--240.

\bibitem{PhysRevE626332}
\BIBentryALTinterwordspacing
V.~N. Belykh, I.~V. Belykh, and M.~Hasler, ``Hierarchy and stability of
  partially synchronous oscillations of diffusively coupled dynamical
  systems,'' \emph{Phys. Rev. E}, vol.~62, pp. 6332--6345, Nov 2000. [Online].
  Available: \url{https://link.aps.org/doi/10.1103/PhysRevE.62.6332}
\BIBentrySTDinterwordspacing

\bibitem{POGROMSKY200265}
\BIBentryALTinterwordspacing
A.~Pogromsky, G.~Santoboni, and H.~Nijmeijer, ``Partial synchronization: from
  symmetry towards stability,'' \emph{Physica D: Nonlinear Phenomena}, vol.
  172, no.~1, pp. 65 -- 87, 2002. [Online]. Available:
  \url{http://www.sciencedirect.com/science/article/pii/S0167278902006541}
\BIBentrySTDinterwordspacing

\bibitem{stewart}
I.~Stewart, M.~Golubitsky, and M.~Pivato, ``Symmetry groupoids and patterns of
  synchrony in coupled cell networks,'' \emph{SIAM Journal on Applied Dynamical
  Systems}, vol.~2, no.~4, pp. 609--646, 2003.

\bibitem{pst}
A.~Y. Pogromsky, ``A partial synchronization theorem,'' \emph{Chaos: An
  Interdisciplinary Journal of Nonlinear Science}, vol.~18, no.~3, p. 037107,
  2008.

\bibitem{Sorrentinoe1501737}
\BIBentryALTinterwordspacing
F.~Sorrentino, L.~M. Pecora, A.~M. Hagerstrom, T.~E. Murphy, and R.~Roy,
  ``Complete characterization of the stability of cluster synchronization in
  complex dynamical networks,'' \emph{Science Advances}, vol.~2, no.~4, 2016.
  [Online]. Available:
  \url{https://advances.sciencemag.org/content/2/4/e1501737}
\BIBentrySTDinterwordspacing

\bibitem{2018arXiv180807263Z}
\BIBentryALTinterwordspacing
J.~Zhang and J.~Zhu, ``Exponential synchronization of the high-dimensional
  {K}uramoto model with identical oscillators under digraphs,''
  \emph{Automatica}, vol. 102, pp. 122 -- 128, 2019. [Online]. Available:
  \url{http://www.sciencedirect.com/science/article/pii/S0005109819300020}
\BIBentrySTDinterwordspacing

\bibitem{jafarpour2018synchronization}
S.~Jafarpour and F.~Bullo, ``Synchronization of {K}uramoto oscillators via
  cutset projections,'' \emph{IEEE Transactions on Automatic Control}, 2018.

\bibitem{ha2016synchronization}
S.-Y. Ha, S.~E. Noh, and J.~Park, ``Synchronization of {K}uramoto oscillators
  with adaptive couplings,'' \emph{SIAM Journal on Applied Dynamical Systems},
  vol.~15, no.~1, pp. 162--194, 2016.

\bibitem{ha2018emergent}
S.-Y. Ha, J.~Lee, Z.~Li, and J.~Park, ``Emergent dynamics of {K}uramoto
  oscillators with adaptive couplings: conservation law and fast learning,''
  \emph{SIAM Journal on Applied Dynamical Systems}, vol.~17, no.~2, pp.
  1560--1588, 2018.

\bibitem{gushchin2016phase}
A.~Gushchin, E.~Mallada, and A.~Tang, ``Phase-coupled oscillators with plastic
  coupling: Synchronization and stability,'' \emph{IEEE Transactions on Network
  Science and Engineering}, vol.~3, no.~4, pp. 240--256, 2016.

\bibitem{bronski2017stability}
J.~C. Bronski, Y.~He, X.~Li, Y.~Liu, D.~R. Sponseller, and S.~Wolbert, ``The
  stability of fixed points for a {K}uramoto model with {H}ebbian
  interactions,'' \emph{Chaos: An Interdisciplinary Journal of Nonlinear
  Science}, vol.~27, no.~5, p. 053110, 2017.

\bibitem{menara2019exact}
T.~Menara, G.~Baggio, D.~S. Bassett, and F.~Pasqualetti, ``Exact and
  approximate stability conditions for cluster synchronization of {K}uramoto
  oscillators,'' in \emph{American Control Conference, Philadelphia, PA, USA},
  2019.

\bibitem{menara2019stability}
T.~Menara, G.~Baggio, D.~Bassett, and F.~Pasqualetti, ``Stability conditions
  for cluster synchronization in networks of heterogeneous {K}uramoto
  oscillators,'' \emph{IEEE Transactions on Control of Network Systems}, 2019.

\bibitem{8263710}
L.~{Tiberi}, C.~{Favaretto}, M.~{Innocenti}, D.~S. {Bassett}, and
  F.~{Pasqualetti}, ``Synchronization patterns in networks of {K}uramoto
  oscillators: A geometric approach for analysis and control,'' in \emph{2017
  IEEE 56th Annual Conference on Decision and Control (CDC)}, Dec 2017, pp.
  481--486.

\bibitem{4961065}
M.~T. Schaub, N.~O'Clery, Y.~N. Billeh, J.-C. Delvenne, R.~Lambiotte, and
  M.~Barahona, ``Graph partitions and cluster synchronization in networks of
  oscillators,'' \emph{Chaos: An Interdisciplinary Journal of Nonlinear
  Science}, vol.~26, no.~9, p. 094821, 2016.

\bibitem{4961435}
I.~V. Belykh, B.~N. Brister, and V.~N. Belykh, ``Bistability of patterns of
  synchrony in {K}uramoto oscillators with inertia,'' \emph{Chaos: An
  Interdisciplinary Journal of Nonlinear Science}, vol.~26, no.~9, p. 094822,
  2016.

\bibitem{5718119}
L.~{Scardovi}, ``Clustering and synchronization in phase models with state
  dependent coupling,'' in \emph{49th IEEE Conference on Decision and Control
  (CDC)}, Dec 2010, pp. 627--632.

\bibitem{berner2018multi}
R.~Berner, E.~Sch{\"o}ll, and S.~Yanchuk, ``Multi-clusters in adaptive
  networks,'' \emph{arXiv preprint arXiv:1809.00573}, 2018.

\bibitem{berner2019self}
R.~Berner, J.~Fialkowski, D.~Kasatkin, V.~Nekorkin, S.~Yanchuk, and
  E.~Sch{\"o}ll, ``Self-similar hierarchical frequency clusters in adaptive
  networks of phase oscillators,'' \emph{arXiv preprint arXiv:1904.06927},
  2019.

\bibitem{PhysRevLett.112.144103}
\BIBentryALTinterwordspacing
A.~Yeldesbay, A.~Pikovsky, and M.~Rosenblum, ``Chimeralike states in an
  ensemble of globally coupled oscillators,'' \emph{Phys. Rev. Lett.}, vol.
  112, p. 144103, Apr 2014. [Online]. Available:
  \url{https://link.aps.org/doi/10.1103/PhysRevLett.112.144103}
\BIBentrySTDinterwordspacing

\bibitem{PhysRevE.96.062211}
\BIBentryALTinterwordspacing
D.~V. Kasatkin, S.~Yanchuk, E.~Sch\"oll, and V.~I. Nekorkin, ``Self-organized
  emergence of multilayer structure and chimera states in dynamical networks
  with adaptive couplings,'' \emph{Phys. Rev. E}, vol.~96, p. 062211, Dec 2017.
  [Online]. Available:
  \url{https://link.aps.org/doi/10.1103/PhysRevE.96.062211}
\BIBentrySTDinterwordspacing

\bibitem{Sam1}
\BIBentryALTinterwordspacing
A.~M. Samoilenko, \emph{Elements of the mathematical theory of multi-frequency
  oscillations}, ser. Mathematics and its Applications (Soviet Series).\hskip
  1em plus 0.5em minus 0.4em\relax Kluwer Academic Publishers Group, Dordrecht,
  1991, vol.~71. [Online]. Available:
  \url{https://doi.org/10.1007/978-94-011-3520-7}
\BIBentrySTDinterwordspacing

\bibitem{mitr}
Y.~A. Mitropolsky, A.~M. Samoilenko, and V.~L. Kulik, \emph{Dichotomies and
  stability in nonautonomous linear systems}, ser. Stability and Control:
  Theory, Methods and Applications.\hskip 1em plus 0.5em minus 0.4em\relax
  Taylor \& Francis, London, 2003, vol.~14.

\bibitem{SAMOILENKO19973121}
\BIBentryALTinterwordspacing
A.~Samoilenko, ``Perturbation theory of smooth invariant tori of dynamical
  systems,'' \emph{Nonlinear Analysis: Theory, Methods and Applications},
  vol.~30, no.~5, pp. 3121 -- 3133, 1997, proceedings of the Second World
  Congress of Nonlinear Analysts. [Online]. Available:
  \url{http://www.sciencedirect.com/science/article/pii/S0362546X96001137}
\BIBentrySTDinterwordspacing

\bibitem{8451900}
F.~{Forni} and R.~{Sepulchre}, ``Differential dissipativity theory for
  dominance analysis,'' \emph{IEEE Transactions on Automatic Control}, vol.~64,
  no.~6, pp. 2340--2351, June 2019.

\bibitem{plasticityHebb}
\BIBentryALTinterwordspacing
P.~Seliger, S.~C. Young, and L.~S. Tsimring, ``Plasticity and learning in a
  network of coupled phase oscillators,'' \emph{Phys. Rev. E}, vol.~65, p.
  041906, Mar 2002. [Online]. Available:
  \url{https://link.aps.org/doi/10.1103/PhysRevE.65.041906}
\BIBentrySTDinterwordspacing

\bibitem{8341829}
L.~V. {Gambuzza}, M.~{Frasca}, and V.~{Latora}, ``Distributed control of
  synchronization of a group of network nodes,'' \emph{IEEE Transactions on
  Automatic Control}, vol.~64, no.~1, pp. 365--372, Jan 2019.

\end{thebibliography}
\bibliographystyle{ieeetran}

\end{document}